\documentclass[journal]{IEEEtran}
\usepackage[utf8]{inputenc}
\usepackage[T1]{fontenc}

\usepackage{amsmath}
\usepackage{bbm}
\usepackage{tabularx,booktabs}
\usepackage{tablefootnote}
\usepackage{capt-of}
\usepackage[caption=false]{subfig}
\usepackage[pdftex]{graphicx}
\usepackage{mathtools}
\usepackage{microtype}
\usepackage{color}

\begin{document}

\title{Going Digital: A Survey on Digitalization and  Large Scale Data Analytics in Healthcare}

\author{Volker Tresp, {J.} Marc Overhage, Markus Bundschus, Shahrooz Rabizadeh, Peter A. Fasching, Shipeng Yu    % <-this % stops a space
\thanks{Volker Tresp is with Siemens AG, Corporate Technology and
    the Ludwig Maximilian University of Munich, Germany.}% <-this % stops a space
    \thanks{Markus Bundschus  is with Roche Diagnostics,  Germany.}% <-this % stops a space
\thanks{{J.} Marc Overhage is with Cerner Corporation, Kansas City, Missouri,  USA.}
\thanks{Shahrooz Rabizadeh is with  NantOmics, LLC and at NantBioScience, Inc,  Culver City, CA, USA.}
\thanks{Peter A. Fasching is with the University of California at Los Angeles, David Geffen School of Medicine, Department of Medicine, Division of Hematology and Oncology, Los Angeles, California, USA and the University Hospital Erlangen, Department of Gynecology and Obstetrics, Erlangen, Germany.}
\thanks{Shipeng Yu is with LinkedIn, Mountain View, California, USA.}}
% \thanks{Manuscript received October 19, 2014; revised January 16, 2015.}}

\maketitle

\begin{abstract}

We provide an overview of the recent trends towards digitalization and large scale data analytics in healthcare. It is expected that these trends   are instrumental in  the dramatic  changes in the way healthcare will be organized in the future. We discuss  the recent political initiatives designed to shift care delivery processes from paper to electronic, with the goals of more effective treatments with better outcomes; cost pressure is a major driver of innovation. We describe newly developed networks of healthcare providers, research organizations and commercial vendors to jointly analyze data for the development of decision support systems.  We  address the trend towards continuous healthcare where health is  monitored by wearable and stationary devices; a related development is that   patients increasingly assume responsibility for their own health data.  Finally we discuss recent initiatives towards a personalized medicine, based on  advances in molecular medicine, data management,  and data analytics.

\end{abstract} 

% \end{document}

%
%% Note that keywords are not normally used for peerreview papers.
%\begin{IEEEkeywords}
%Healthcare, Personalized Medicine, Precision Medicine, Big Data, EHR,
%\end{IEEEkeywords}

%
%\begin{table}[!t]
%% increase table row spacing, adjust to taste
%\renewcommand{\arraystretch}{1.3}
%%  if using array.sty, it might be a good idea to tweak the value of
%%  \extrarowheight as needed to properly center the text within the cells
%\caption{An Example of a Table}
%\label{table_example}
%\centering
%% Some packages, such as MDW tools, offer better commands for making tables
%% than the plain LaTeX2e tabular which is used here.
%\begin{tabular}{|l l l |}
%\hline
%HITECH Act  & Health Information Technology for Economic and Clinical Health Act & x\\
%\hline
%EHR & Electronic Healthcare REcord & Five\\
%\hline
%\end{tabular}
%\end{table}

\section{Introduction}
\label{Introduction}

% surrogate
% generate hypothesis

% healthcare provider .... caregiver

Data \textcolor{black}{have} always been the basis for a scientific approach to healthcare: diagnostics are supported by physiological measurement,  laboratory data and diagnostic imaging;
\textcolor{black}{the analysis of treatment efficiency  and  potential disease causes is based on clinical and epidemiological studies.}
Study design and data acquisition used to be the main challenges whereas data volume and data management were not. We expect that this will change rapidly as new sources of healthcare data become increasingly relevant.
The \textcolor{black}{newly} generated  data sets
 are high-dimensional and abundant;  \textcolor{black}{data volume} is simply exploding.
 In the same sense as ``digitalization'' stands for  the \textcolor{black}{increasing digital presence of individuals},  services, and  ``things'' in general, the term ``digital health'' is associated with the wealth of health-related data becoming available in digital form~\cite{biesdorf2014healthcare}.
 The management and the analysis of these data with the goal of gaining insights and making insights actionable is sometimes referred to as Healthcare Big Data~\cite{manyika2011big,conger2012data,kayyali2013big}.
 Whereas the  term ``Big Data'' might quickly fall out off fashion, the underlying issues and technological challenges covered in this paper  most likely will  not.

Driving forces for the changes include a number of  recent political initiatives designed to shift care delivery processes from paper to electronic, with the goals of more effective treatments with better outcomes; cost pressure is a major driver of innovation.
One example is the Health Information Technology for Economic and Clinical Health Act (HITECH Act) in the U.S.
The focus of the HITECH Act is the meaningful use of an interoperable electronic health record (EHR), enabling the exchange of information across institutions.
 The overriding goals are that each involved healthcare professional \textcolor{black}{has} complete patient information, that patients are treated by the best available institution for their problems,  that medical research results can have more immediate impact, and that overall effectiveness is increased.
 In the context of these initiatives, large volumes of data will be collected and many improvements in healthcare will be based on the analysis of \textcolor{black}{these data},  with improved outcome at manageable cost as main goal. As a precondition to realizing the full potential,  fundamental changes in the healthcare system might be required and data privacy, data ownership and data security issues must be resolved.

 ``Variety'' and ``volume'' are the Big Data aspects most relevant to  healthcare.
Variety  means that   detailed information  about an individual must be  available to personalize recommendations and interventions. Examples \textcolor{black}{of the latter two} are    lifestyle recommendations, alarms, reminders, preventive measures, screenings,   referrals, and treatment recommendations.
 Key issues are, first,  how detailed patient information can be acquired, managed and stored, second, how the ``intelligence'' comes into the system and, third,  how recommendations should be optimally communicated to stake holders.

Volume is important to gain valid insights and actionable solutions from \textcolor{black}{healthcare data: If} data on many individuals \textcolor{black}{are} collected, one can perform statistical analysis, data mining and train machine learning \textcolor{black}{algorithms}.

The goal of this paper is to provide an overview of how digital health \textcolor{black}{might  affect} the future of healthcare --- and the expected changes are dramatic.
The paper is written for the interested reader with limited prior exposure to healthcare issues.
It contains six major sections ---organized along the digitalization  sources---  which describe different digitalization and analytics trends in some detail.

In the next section we consider the digitalization process within the clinic.
As mentioned,  many advances in clinical data management  are based on   a  broader adoption of the EHR, which is the main driver for a digitalization of the clinical information systems.
\textcolor{black}{The introduction of a high quality} EHR can lead to an  improvement in  patient safety and can increase  transparency and accountability.
It documents relevant clinical patient information and \textcolor{black}{its data are} the basis for many forms of analysis and decision support.
 Implementing EHRs faces challenges,  mostly associated with the additional efforts and costs and the fear that the center of attention might move away from the patient to the IT system.
   We also  discuss  the current clinical data situation:  what type of data  \textcolor{black}{typically are} available and how \textcolor{black}{they are}  documented and organized.
     \textcolor{black}{We discuss the importance of shared terminologies, data security and data privacy.}

Clinics increasingly collaborate  in digitalization and Big Data projects with \textcolor{black}{university institutes with analytics} competencies  and with commercial vendors.
In Section~\ref{Mobilizing} we describe a few specific  projects.
\textcolor{black}{ We also discuss some of the statistical issues that arise in integrating observed data from different sources, e.g., various  biases, hidden confounders and batch effects.}

 % Another important issue is  the representation of the temporal aspects of patient data.

Payers, registries and national health systems, as the U.K.'s NHS,  have long collected healthcare related data across clinics.  A novel development in recent years is that clinics are increasingly required to report data for purposes like quality control and policy development.
\textcolor{black}{Also, a lot} can be gained if data can be exchanged between different care venues, as for example in integrated care.
{Health information exchange (HIE) encompasses all activities towards a mobilization of digital  healthcare information
 across organizations within a region, community or hospital system~\cite{wikiHIE}.}
\textcolor{black}{We discuss the U.S. HIPAA regulations and the  danger  of data de-identification}.
The externalization of clinical data is  covered in Section~\ref{External}.

Healthcare \textcolor{black}{increasingly becomes} patient-centered and  patients want to get in charge of their own health and their own health data.
Families want to keep health profiles and make them accessible to authorized caregivers, like their family doctors.
 \textcolor{black}{These trends are} supported by a number of evolving cloud-based offerings.
 One can envision new IT platforms as basis for a revolution in healthcare management,  supporting both a patient centric and a data centric view.
Also, there are patients with one or several serious, sometimes chronic  diseases who want to interact with a social community of patients with similar problems.
 Social media used by these patients  may provide insights into drug effectiveness, adverse drug effects and can be useful for the detection and the tracking of infectious \textcolor{black}{diseases}.
  Patients are sometimes willing to make their data available  for research and other uses via platforms like PatientsLikeMe.
   We discuss these developments in Section~\ref{Charge}.

Another big digitalization trend is increasing data capture during the course of everyday activities. Smart phones can collect  fitness and health related data \textcolor{black}{via a variety of} sensors.
\textcolor{black}{These} data can be analyzed by  patients via platforms and apps  \textcolor{black}{and they} can be communicated to  \textcolor{black}{healthcare providers.}
 Over a patient's lifetime large amounts of personal data \textcolor{black}{are} collected and data analytics is  offered as a service by platform operators.
\textcolor{black}{
 This mobile health (mHealth) supports efforts to ``shift care to the left'', i.e.,  to identify risk and intervene before disease develops; there is an increasing emphasis on prevention,  rather then diagnosis and treatment. In addition to wearable devices, ambient sensors will play a role, in particular for the care of the elderly.}
%This trend often manifests itself in the broader application of predictive models and highly targeted testing to individuals and an increasing emphasis on prevention rather than diagnosis and treatment.
%
These developments are discussed in Section~\ref{Continuous}.

Finally,  there is a growing  trend towards personalization in healthcare (i.e., more precise and personalized care) partially but not solely driven by the lower cost and increasing availability of molecular  data in form of  genomic (including \textcolor{black}{the} whole genome), proteomic  and \textcolor{black}{metabolic profiles.
%The analysis of germline DNA characterizes disease predispositions whereas the analysis of somatic cancer DNA and RNA  characterizes the disease.
Treatment decisions are more and more }  based \textcolor{black}{on molecular} patient profiles; as a drawback, \textcolor{black}{personalization}   comes at an increased level of complexity that easily overwhelms the decision maker.
Large-scale analytics is essential for the generation of    personalized decision rules  derived from large sets of data, in line with the trend towards an evidence-based medicine.
\textcolor{black}{These are the topics} of Section~\ref{Personal}.

Section~\ref{Conclusion} summarizes the developments  with an attempt of an evaluation and  a discussion of opportunities and  challenges.
In this paper we focus primarily on the situation in the U.S. The main reason is that the U.S. is ahead in digitalization and large scale \textcolor{black}{data analytics in general\textcolor{black}{,}  and in healthcare in particular. }
 Another reason is that the U.S. has the largest healthcare market worldwide. We will highlight the situation in other countries when relevant;  in particular some of the developments in the U.K. are highly innovative and demonstrate emerging opportunities in  a national healthcare system.

%\textsc{Overriding issues in  all developments are  data privacy, data security, viable business and reimbursement models,   and legal certainty.}

%
  \section{Digitizing Healthcare Data}
\label{dcd}

\subsection{Motivation}

Healthcare is a large and complex enterprise that is relevant to  every person on the planet.
The digitization of  healthcare data in a manner that is easy for computers to utilize is important to support the delivery of care through data visualization, collaboration and clinical decision support.
 Recently, the concept of a ``learning healthcare system'' has been introduced~\cite{lhcs2012}.
In a learning healthcare system data harvested from the care process is continuously analyzed and used to create insights into how the care delivery process should evolve.

When data \textcolor{black}{are} digitized,  \textcolor{black}{it is possible to  create new and useful ways
for visualization and analysis,}  with the potential to provide better insights into a patient's status and, optimistically, better decisions~\cite{rind2011interactive}.
Another important application for digitized healthcare data is \textcolor{black}{digitally supported} clinical decision support (CDS).  CDS systems combine the data with clinical knowledge to provide patient specific suggestions at the appropriate time in the care process.
These systems have been demonstrated to improve the quality, safety and efficiency of care,  though these advantages  have not been universally observed~\cite{musen2014clinical,bright2012effect}.
Lack of complete, timely and correct data frequently underlies the failure to achieve these benefits.

Complete information  from \textit{many patients} is the basis for analytics, i.e., statistical analysis, data mining  and machine learning.
A few authors have attempted to characterize the ways that healthcare systems hope to take advantage of analytics.  Bresnick and colleagues  considered  the following items~\cite{Bresnick2015-2}:
\begin{itemize}
  \item 	Identifying at-risk patients
  \item 	Tracking clinical outcomes
  \item 	Performance measurement and management
  \item 	Clinical decision making at the point of care
  \item 	Length of stay prediction
  \item 	Hospital readmission prediction
\end{itemize}
The goal of the latter is to avoid  costly penalties for hospital readmissions, which   were introduced by Medicare under the 2010 Patient Protection and Affordable Care Act (ACA).
Insurance companies have started to use data analytics to identify likely \textcolor{black}{patients for hospital readmissions, which resulted in a  40-50\% reduction for patients} with  congestive heart failure~\cite{Blumenthal2015,Bresnick2015-2}.

Another study identifies the following uses for analytics~\cite{SAS2015}:
\begin{itemize}
  \item Analytics-based drug discovery processes;  study of drug efficacy; detection of adverse drug effects and drug-drug interactions
  \item Identification of  better and safer therapies
  \item Optimal clinical trial designs and patient recruitment
  \item Evidence-based medicine to integrate clinical expertise and research results to support best \textcolor{black}{care decisions}
  \item Protocol-based medicine that draws on research results to identify best practices for specific conditions, medical histories and patient populations
  \item Personalized medicine that blends diverse data sources, including genetic profiles, with historical clinical data
\end{itemize}

These are  mixes of descriptive tasks, prediction tasks and prescriptive tasks~\cite{Bresnick2015-2}.

\textit{Descriptive analytics} is a classical data mining task and extracts  human understandable   information  from data  in form of simple rules (association rule  mining) or  in visual form (visual analytics)~\cite{DBLP:books/crc/p/Hersh15}. Often the results are presented as a report. Typical projects might be to  identify areas for improvement on clinical quality measures or on specific aspects of care. It is important to note that the human is in the loop and draws conclusions based on the findings~\cite{Bresnick2015-2}.

For \textit{predictive analytics}, traditional statistical methods or machine learning can be used.
The task might be to forecast  future procedures, diagnoses, or outcomes.
Other tasks are patient  condition monitoring with different alarm functions.
The application of predictive models \textcolor{black}{at the point of care   requires} a robust and high-quality infrastructure, which enables real-time data processing.
``Medical devices must be fully integrated to provide up-to-the-second information on patient vitals to improve safety, while alerts and alarms have to be developed and presented to clinicians without hopelessly disrupting their workflows or annoying them into ignoring critical warning''~\cite{Bresnick2015-2}.
The good news is that confounding factors, as long as their statistical properties are stationary, can be ignored in pure prediction problems; on the other hand,  a predictive model trained in one clinic might not work well in another clinic, e.g., due to different patient profiles.

\textit{Prescriptive analytics} encompasses the ability to recommend actions and to answer ``what if'' type of  questions.
Whereas a predictive model might recommend an action that is ``typically''  performed for a patient with particular properties, a prescriptive analysis would be able to prescribe an action that would lead to best \textcolor{black}{predicted} outcome.
``Prescriptive analytics doesn’t just predict what’s likely to happen, but actively suggests how organizations can best take action to avoid or mitigate a negative circumstance''~\cite{Bresnick2015-2}.
The requirements on data quality and system robustness are even greater.
 In particular, a prescriptive analysis requires a careful analysis and consideration  of hidden confounders.
Prescriptive analytics has been called ``the future of healthcare Big Data \ldots  the healthcare industry has an enormous opportunity \textcolor{black}{by taking} advantage of these decision-making abilities''~\cite{Bresnick2015-2}.

\subsection{The Electronic Health Record}

For decades, much of what  was documented about a patient was in paper format and collected in a folder that was physically moved across the clinical departments and was eventually filed.
Today, patient data \textcolor{black}{are} increasingly recorded and stored in an electronic form, the electronic health record (EHR)~\cite{DBLP:books/crc/p/RahmanR15}.
 The EHR  greatly improves the quality of the data documented and
  supports improvements in  patient care by enabling analysis and decision support.
 In its most basic form, an EHR consists of the same paper documents except that they are scanned and stored digitally. Of course this does little to support analysis or clinical decision making.
More advanced systems contain machine readable structured tables and digital reports, where ideally  the latter are machine \textcolor{black}{readable   and} semantically annotated.
 In these advanced systems,  data \textcolor{black}{are}   easily accessible to algorithms and analytic tools.  \textcolor{black}{}

%  A number of interactive EHR visualization tools have been developed to  get insights into  its content~\cite{west2015innovative}.
%

 As we will discuss in Section~\ref{External}, the HITECH Act has stimulated increased use of the EHR in both hospitals and ambulatory practices across the U.S.
~\cite{blumenthal2010launching,charles2014adoption,hsiao2012use}.
\textcolor{black}{ Meaningful use, as defined by HITECH, requires both the capability and actual use of the EHR to perform functions such as
 electronic prescribing and ordering of tests, electronic access to test results, medication alerts, and tracking of lab tests. In addition  medical guideline support must be implemented.}
  In some countries, the EHR is standard (e.g., in the Netherlands, New Zealand, Norway, Sweden and the U.K.), whereas countries like the U.S. and Germany are lagging behind.   Surveys found that, despite much broader adoption over the last several years,  U.S.  physician enthusiasm for EHRs has not improved in the last 5 years~\cite{McCarthy2015}.
The authors attribute the physician's lack of enthusiasm to doctors not seeing enough benefit from the EHR and that EHR products do not deliver all necessary functionalities, being difficult to use,  and not being  interoperable with each other.
In addition, there are worries about data leakage, which is increasing in frequency~\cite{snell2015}, and compliance with regulations.

\subsection{Structured Data Capture}

EHRs can only achieve  their full potential if time and cost associated with data capture can be kept under control.
 While a good deal of clinical data can be obtained from other venues  such as laboratory or radiology systems or from devices (e.g. vital signs, ventilators),  a significant amount of data must be entered by providers.
 Because of the time and effort required for providers to capture structured data, they often question if there is sufficient value to warrant the negative impact on productivity~\cite{clynch2015medical, friedberg2013factors}.
 Contemporary EHRs are estimated to require an additional 48 minutes per day,  much of which is devoted to documentation~\cite{mcdonald2014use,mcdonald2012invited}.

% \subsection{Coding and Coding Systems}

Healthcare is complex,  which is also  reflected in the data:  There are hundreds of thousands of clinical concepts that have to be represented.
In order to accommodate this scale and simplify representations, coding systems have been adopted for clinical concepts.
The concept of heart failure for example can be represented in the International Classification of Disease Version~9 Clinical Modification as "428.0".
%  This approach facilitates using \textcolor{black}{key-value} approaches to representing data.
 Unfortunately, there are multiple coding systems for most clinical concepts,  so heart failure can also be represented by I50 (ICD-10), 16209 (DiseaseDB), D00633 (MESH), 42343007 (SNOMED) and others.
 Even more unfortunately, a good deal of data are coded using idiosyncratic clinical codes that are unique to a specific healthcare delivery system.
  This variation means that using the data often requires mapping or translation between coding systems which usually requires substantial human effort and, in some cases, a specific data model.

 In addition to direct entry by providers or their surrogates, structured data can be derived directly  from unstructured data including free text, images and other signals.

% images
Radiology involves the acquisition, analysis, storage and handling of radiological images and certainly involves huge amounts of data, in particular when the analysis involves time, as in angiography, or all three spatial dimensions, as in whole body screening.
Pathology involves the analysis of tissue, cell, and body fluid samples, typically via microscopic imaging. As pathology is digitized, increasing amounts of digital data \textcolor{black}{are} generated and need to be handled and stored.
 The standard is that  medical specialists interpret the radiological and pathological  images and describe the  findings in  written  free-text or unstructured reports, although there is a trend towards template-based semi-structured reporting.
% Increasingly,  algorithms can be applied to extract structured data directly from the images themselves.

The computerized analysis of radiological and pathological images is an established research area involving sophisticated algorithms and is becoming increasingly clinically relevant~\cite{mihalef2011patient,DBLP:books/crc/p/PadfieldMG15,DBLP:books/crc/p/LiaoYWHZSPZBS15}.
The analysis typically involves some form of machine learning and the emerging field of deep learning has increasing impact~\cite{enlitic2015}. \textcolor{black}{Analysis generates} qualitative and quantitative labels or tabs, which can be used in integrated analytics studies~\cite{tresp2013towards}.

% text
Written text is a major medium:  The exact numbers vary, but a significant proportion of the clinically relevant information is only documented in textual format.
Besides radiological and pathological reports,  medically relevant  textual sources are  reports from other departments, notes, referral letters  and discharge letters.
Both researchers and commercial developers have devoted considerable effort to improve the efficiency of structured data capture from text and some hope that Natural Language Processing (NLP) will obviate the need for structured data capture;  but advances  have been incremental;
while there is progress in focused areas, information extraction  from clinical texts is notoriously difficult.
Some of the reasons are that reports are ungrammatical, contain short phrases, non-standardized and overloaded abbreviations and employ an abundant use of negations and lists.
Structured reporting, where the text is generated automatically and the physician simply enters keywords and short pieces of text, would be a great advance, but is currently not the standard~\cite{DBLP:books/crc/p/RajaJ15}, in part because it is typically more time consuming for the provider.

Another issue is that the structured data entered by providers or extracted from text \textcolor{black}{need} to be represented such that \textcolor{black}{they } can be ``understood'' by a computer, in other words \textcolor{black}{healthcare systems need} to be able to communicate effectively and in the same formalized language.
Some languages are essentially simple taxonomies and vocabularies and are the basis for standards used in the billing process, such as ICD for diagnosis, CPT\copyright for procedures, and SNOMED codes for diseases or conditions.
For medications, there is the National Library of Medicine's   RxNorm, the National Drug Code (NDC) and others.
 Logical Observation Identifiers Names and Codes (LOINC\copyright) define universal standards for identifying medical laboratory and clinical observations.
 
 For billing purposes  all involved players are highly motivated  to employ the codes with great discipline.
 Implied statements in general take on simple forms, like ``Patient X has Disease Y''.

This changes if one wants to express some detailed medical finding accurately.
Consider the phrases ``43 yo female with history of GERD woke up w/ SOB and LUE discomfort 1 day PTA. She presented to [**Hospital2 72**] where she was ruled out for MI by enzymes. She underwent stress test the following day at [**Hospital2 72**]. She developed SOB and shoulder pain during the test.''
In order to utilize the information  represented in this text,
  an application would first need to map and code the entities in the phrases and then formulate statements relating the complex sequential observations with many subtle phrases \textcolor{black}{only understandable by  trained experts.} \textcolor{black}{These challenges} goes far  beyond the expressiveness of currently used  medical formal languages.

%   In addition, we would need  to generate simplified versions of the statement which  can easily be queried.

Genomic, proteomic and other molecular data (discussed more fully in Section~\ref{Personal}), which are almost by their nature digital, will add an extensive amount and variety of structured data though,  in current practice, an extremely limited subset  derived from the molecular data will be all that is necessary for a particular application.

\subsection{Data Silos}

Other barriers to utilizing clinical data are the ubiquitous clinical data silos.
In addition to the fragmentation of a patient's data across various participants in the healthcare ecosystem,  each  medical department   historically has used its own department-specific database and reporting system,  and only a portion of that information has \textcolor{black}{typically} been integrated into the EHR~\cite{mcdonald1998canopy}.
As an example, before a provider sees a laboratory test result displayed in their EHR, the data \textcolor{black}{have} traveled along a complex and convoluted path to get there: Laboratory instruments themselves are sophisticated computing and data management systems that  pass data through laboratory instrument management systems and potentially laboratory information systems, through an interface engine and eventually to the EHR.
Each phase supports specific data management and monitoring tasks and adds and loses pieces of data~\cite{Bresnick2015}.
Another issue is that each data silo might code information differently,  and building wrappers for the purpose of data integration is anything but simple.
These challenges are the basis for the recent preference for integrated EHR platforms \textcolor{black}{to} share a common database across many departments, which largely eliminate the data silos inside an organization.
In fact healthcare organizations have often accepted lesser functionality in order to achieve this benefit.

%
%Figure~\ref{clinic} shows some of the clinical data silos and a typical patient story.
%
%\begin{figure}
%  \centering
%  % Requires \usepackage{graphicx}
%  \includegraphics[width=\columnwidth]{clinic.PNG}\\
%  \caption{
%  Consider a patient that is referred to a specialty clinic from a medical practitioner for management of a certain problem. In the example in the figure, we  assume that this is a nephrology patient with kidney disease being cared from by a nephrologist. In a clinic there are also many other departments like cardiology, neurology, urology, gynecology, and so on.   When the patient arrives at the clinic, administrative data will be recorded, like name, address, payer and employer.  There might be a referral request or transfer letter from the medical practitioner stating the problem and the treatment status.  The attending physician will perform an assessment and an initial diagnosis might be documented.  Diagnostic tests will be performed to confirm the diagnosis or severity and a treatment might be agreed upon.  The patient leaves the clinic with a discharge letter summarizing diagnosis, treatments and prognosis. Further follow-up treatments might be required,  which might be summarized in the discharge letter.
%  }
%  \label{clinic}
%\end{figure}
%

\subsection{Clinical Data Integration Efforts}

Some providers may have implemented a separate research data system such as i2b2~\cite{murphy2010serving} or tranSMART~\cite{athey2013transmart}.
These systems extract clinically relevant information from the EHR and from other clinical resources and databases and integrate them into the research database.
A research database can be a great resource for \textcolor{black}{data analytics projects}.
Unfortunately installing a research database  can be extremely demanding since  it needs to access data \textcolor{black}{from} the data silos of the different departments.
 As discussed these databases might all have different structures and use different terminologies.

In contrast to clinical data, billing data ---in part because of its simplicity and in part out of necessity--- \textcolor{black}{are} consistently structured and \textcolor{black}{are} often part of a research database.
Unfortunately, \textcolor{black}{billing data} does not contain much of the clinically relevant information and may not accurately and fully reflect clinical reality. \textcolor{black}{Reasons are that providers} may not be as careful in recording administrative data believing  \textcolor{black}{that it is} not critical to be exactly correct or, in some cases, billing data may \textcolor{black}{be} coded to maximize reimbursement rather than to most accurately reflect the patient's clinical status.

Another important issue is that the temporal order of events is often not well  documented in the data. To analyze the causal effects of a decision and to optimize decisions, it is important to  know which information was available to the decision maker at the time of decision. At the current status of documentation,  reconstructing the temporal order of events  can be difficult.

\subsection{Privacy Protection and De-identification}

De-identification is the process used to prevent a person’s identity from being connected with information.
 Common uses of de-identification include human subject research,  which requires privacy protection for research participants.
 Common strategies for de-identifying data sets are deleting or masking personal identifiers, such as name and Social Security Number, and suppressing or generalizing quasi-identifiers, such as date of birth and ZIP code.
  More sophisticated approaches use k-anonymity, l-diversity, epsilon differential privacy, differential identifiability coarsening, imputation, and data swapping~\cite{DBLP:books/crc/p/ParkG15}. \textcolor{black}{Unfortunately,  information can be lost in de-identification,  making the data potentially less useful for analysis. }

De-identification is difficult for clinical data in general  but particular difficult for textual data since  a personal identifier  might appear unexpectedly in the middle of a text and \textcolor{black}{also} for genomic data, considering that a person's genetic profile is unique.

Appropriate patient consent may reduce the need for de-identification~\cite{danezis2015privacy}.

  \section{Mobilizing Data in a Trusted Network}
\label{Mobilizing}

Integrated care is a worldwide trend in healthcare with the goal of achieving  a  more coordinated and integrated form of care provision.
It may be seen as a response to the problems associated with the fragmented delivery of health in many countries.
Integrated care ---as some other forms of alliances and inter-clinical collaborations---  permits the integration and evaluation of data from several sources. It  supports analytics projects since  the patient sample size simply is larger if compared to  a single clinic, and since patients may stay for more problems   within an integrated care system and for a longer time span, possibly all their life;  thus data on a particular individual \textcolor{black}{are typically} more complete.

In this section we describe  representative projects where  clinic networks team up  with research centers ---which provide expertise in data analytics, machine learning, and medical informatics--- to explore the potential of clinical data analytics.
The long-term  vision behind these and similar projects is a system where patient data \textcolor{black}{are} analyzed online, and research insights   rapidly becomes  common practice, resulting in best care for each patient.

\subsection{The Pittsburgh Health Data Alliance}

The Pittsburgh Health Data Alliance is a collaborative Big Data effort involving Carnegie Mellon University (CMU), the University of Pittsburgh (Pitt) and the University of Pittsburgh Medical Center (UPMC). It is financed by the latter but all three institutions contribute grant funding~\cite{raths2015}.

{The stated goals are characteristic for these types of  projects:
Primarily \textcolor{black}{the consortium seeks}  to analyze and make use of the massive \textcolor{black}{amounts} of data generated in  \textcolor{black}{the}  healthcare system, including EHR patient information, diagnostic imaging, prescriptions, genomic profiles, insurance records, and data from wearable devices.
The \textcolor{black}{work will} support the development of evidence-based medicine, and lead to the augmentation of disease-centered models with patient-centered models of care.
The vision is a  data-driven medicine based on a  large sample of  patients, which  will assess  an individual’s disease risk and make personalized recommendations for treatments.  Other intended outcomes are spinoff companies and promotion of  economic development in the region~\cite{cmu-spyce}.}

The CMU plans to develop  an automated patient diagnosis system. Based on automatically retrieved  symptoms and lab findings  the system searches medical literature and analyzes patient data to provide possible diagnoses. To refine the diagnosis additional tests might be requested.

The role of Pitt's Center for Commercial Applications of Healthcare Data (CCA) is to develop  new technology for potential use in commercial theranostics, combining diagnostics with therapy and imaging systems. UPMC Enterprises  leads the efforts to transfer the results  to   for-profit startup companies. A concrete collaboration \textcolor{black}{topic} concerns the early detection of disease outbreaks by tracking of over-the-counter medication sales. Involved are  the   ``Real-Time Outbreak of Disease Surveillance'' (RODS) Laboratory at Pitt and the  ``Event and Pattern Detection'' (EPD) Lab at CMU’s Heinz College.

Being one of the first sizeable  Big Data projects in  healthcare, the effort attracted the interest of a number of IT companies, which are supplying high-performance database platforms, business intelligence solutions, \textcolor{black}{and  platforms} for  integrating patient records.
In general,  there is an increasing care provider  demand for  Big Data functionalities in clinical information systems and vendors \textcolor{black}{are adapting} to these needs.
In fact, considering the dramatic changes expected in \textcolor{black}{healthcare,   in which} IT is expected to play a major role, many IT vendors are actively exploring future business opportunities.

\subsection{The Mayo Project}

A collaborative effort between \textcolor{black}{the} Mayo Clinic and several departments at the University of Illinois is part of a large federal grant for the support of medical Big Data research~\cite{Flanagan2014}.
The collaborative effort involves the Institute for Genomic Biology, the Department of Computer Science, the Coordinated Science Laboratory, the College of Engineering and the National Center for Supercomputing Applications (NCSA).
The effort includes  the setup of a  new Center of Excellence for Big Data Computing and a network to move and share the data between researchers.
The  Campus Advanced Research Network Environment (CARNE) has been created with the goal of providing unrestricted high-speed access to off-campus locations for specific research purposes.
A major project is the  Knowledge Engine for Genomics, or \textit{KnowEnG}\footnote{http://www.knoweng.org/}.

\subsection{Neonatal Intensive Care at Kaiser Permanente}

This is an early project that    demonstrated the potential  of Big Data in intensive care.
In current medical practice, newborns are typically taken to the neonatal intensive care unit (NICU) if the mother’s temperature rises above a threshold because this may signal an increased risk of neonatal sepsis, a bacterial blood infection~\cite{marcum-2014}.
Kaiser Permanente has used data analytics to develop the interactive and online
 ``Newborn Sepsis Calculator''
  that determines the probability of neonatal sepsis, allowing the care team to better determine which babies to evaluate and treat for infection~\cite{Byron2014}.

\subsection{Indiana Network for Patient Care}

The Regenstrief Institute was an early advocate for clinical data interoperability based on information standards and leveraged that work to enable health information exchange both regionally and nationally.
Regenstrief investigators implemented the Indianapolis Network for Patient Care (INPC) in 1995 with the goal of providing clinicians with data necessary for patient diagnosis and treatment at the point of care.
In 2016, over 100 hospitals, thousands of physician practices, ambulance services,  large local and the state public health departments, regional laboratories and imaging centers, and \textcolor{black}{payers} participate in the INPC.
The federated data repository stores more than 4.7 billion records, including over 118 million text reports from almost 15 million unique patients.
The data \textcolor{black}{are} stored in a standard format, with standardized demographic codes;  laboratory test results are mapped to a set of common test codes with standard units of measure;  medications, diagnoses, imaging studies, report types are also mapped to standard terminologies.
The flows of data\textcolor{black}{, which  enable the INPC,}  support results delivery, public health surveillance, results retrieval, quality improvement, research and other services.
Building on this experience, Regenstrief investigators have informed the development of the nationwide health information network program now called the eHealth Exchange (``Exchange'').

The INPC data \textcolor{black}{have} been utilized by Regenstrief for many Big Data studies and projects including:
\begin{itemize}
  \item 	The OMOP (Observational Medical Outcomes Partnership)~\cite{overhage2013desideratum}   and the subsequent OHDSI (Observational Health Data Science and Informatics)~\cite{hripcsak2015observational} projects to utilize large scale observational data for  drug safety studies
       \item The two projects were a  basis for   ConvergeHEALTH, an effort spearheaded by Deloitte that aims to offer comprehensive data sharing among key organizations. Deloitte has an analytics platform that allows hospital systems to compare results with tools designed to study certain patient outcomes: their OutcomesMiner tool helps users explore real-world outcomes for sub-populations of interest
  \item The Merck-Regenstrief Institute ``Big Data'' Partnership – Academic-Industry Collaboration to Support Personalized Medicine was formed in 2012 to leverage the INPC to support a range of research studies that use clinical data to inform personalized healthcare. The partnership has funded 50 projects to date. Industry commentators have observed that such partnerships between industry and academia, and between and among other payers, are essential as neither sector alone can undertake such projects
\item The Indiana Health Information, a non-profit organization created to sustain the INPC’s operations\textcolor{black}{,}  entered into a partnership agreement with a commercial predictive analytics company, Predixion, to develop new predictive applications aimed at further supporting the patient and business needs of ACOs and hospitals. The INPC database supports Predixion’s current and future solution development
\end{itemize}

\subsection{Clinical Data Intelligence}

Clinical Data Intelligence (“Klinische Datenintelligenz”) is a German project funded by the
\textcolor{black}{Federal Ministry for Economic Affairs and Energy (BMWi) }
and involves two integrated care providers, i.e., the University Hospital Erlangen and the \textcolor{black}{Charit\'{e}} Berlin, two globally acting companies, i.e., Siemens AG and the Siemens Healthineers,
and application and research centers from the University of Erlangen, the German Research Centre for Artificial Intelligence (DFKI), Fraunhofer, and Averbis~\cite{tresp2013towards,sonntag2015clinical}.

 The project puts particular emphasis on terminologies and ontologies, on metadata extraction from textual sources and radiological images and on the integration of medical guidelines as a form of prior knowledge.
 As part of the project a central research database is installed which serves all research and application subprojects.
 The project also addresses business models and \textcolor{black}{clinical} app infrastructures suitable for large-scale data analytics.

The core functionalities  are realized by an integrated learning and decision system (ILDS).
The ILDS  accesses all patient specific data and provides analytics, predictive and prescriptive functionalities.
 The ILDS models and analyzes clinical decision processes by learning from the EHR's structured data  such as diagnosis, procedures, and  lab results.
 The ILDS also analyzes medical history,  radiology, and pathology reports
  and includes guideline information.
\textcolor{black}{  In addition, the ILDS  considers  genomic data, and molecular data in general, to explore the application of personalized medicine to clinical practice.}

The ILDS will immediately be able to make predictions about common practice of the form: ``For a patient with properties and problems X,  procedure Y is typically done (in your clinic system)''.  More difficult, since it involves a careful analysis of confounders, is a prescription of the form: ``For a patient with  with properties and problems X, procedure Y is typically done (in your clinic system) but procedure Z will probably result in a better outcome''.

An important outcome of the project will be a set of requirements for a \textcolor{black}{future} clinical documentation that will enable more powerful data analytics in the future.
\textcolor{black}{For example, patient complaints, symptoms, and clinical outcome are} not always well documented. Readmission within a certain period of time (typically a month) is sometimes taken for a negative outcome.
Alternatively one might define a hospital stay of more than a certain number of days as a negative outcome, where the threshold is  specific to the Diagnosis Related Group (DRG).
In some cases, for example after a kidney transplantation or mastectomy, the patient is closely observed, and outcome information is available\textcolor{black}{,} possibly over patient lifetime.

The ILDS partially uses  deep learning (more specifically recurrent neural networks) to model the sequential decision processes in clinics~\cite{esteban2015}.\footnote{Deep Learning is one of the most exciting developments in machine learning in recent years. It is a field that attracts amazing talents with stunning successes in a number of applications. One of the driving forces in Deep Learning  is DeepMind, a London based company owned by Google.
DeepMind Health  is a project in which \textcolor{black}{U.K. NHS's} medical data \textcolor{black}{are} analyzed.
The agreement gives DeepMind access to  healthcare data \textcolor{black}{of} more than a  million patients~\cite{dmh2016}.
A first outcome is the mobile app Streams, which presents timely information that helps nurses and doctors detect cases of acute kidney injury.
% By comparing  patients’ information with millions of other cases,  the app might be able to predict that they are in the early stages of a disease that has not yet become symptomatic. Additional test can then be run to determine
% if the prediction is correct.
Other  notable commercial Deep Learning efforts with relevance to healthcare  are
Deep Genomics\footnote{http://www.deepgenomics.com/},
Entlitic \footnote{http://www.enlitic.com/} and
Atomwise\footnote{http://www.atomwise.com/}.}

%Considering the overwhelming success of Deep Learning in image analysis, one can expect that many companies will follow. One of the first approaches for using Deep Learning (more specifically recurrent neural networks) to model the sequential decision processes in clinics is described in~\cite{esteban2015}.
%
%\textsc{AtomNet: A Deep Convolutional Neural Network for Bioactivity Prediction in Structure-based Drug Discovery}

The project addresses two use cases in detail.

The first concerns nephrology.
Kidney diseases cause a significant financial burden for the healthcare system.
The aim of this work is to systematically investigate drug-drug interaction (DDI) and adverse drug reactions (ADR) in patients after renal transplantation and to realize an integrated decision support system.
The use case is particularly interesting since longitudinal data covering several decades are available \textcolor{black}{and since outcome information is available.}

First ILDS results are reported in~\cite{esteban2015}, \cite{esteban2016predicting}.

The second use case concerns breast cancer, the most common malignancy in women. Relevant events are screening, diagnosis, therapy and follow-up care.
Of special interest here is the determination of risk factors, the evaluation of the therapy and the prediction of side effects.

% >>>>
% The integration of genome information in clinical decision support is sometimes referred to a clinicogenomics.

\subsection{Related Initiatives and Projects}

In the U.S. and in other countries many similar initiatives have been started or are in preparation phase.

The Dartmouth Institute, Dartmouth-Hitchcock, Denver Health, Intermountain Healthcare, and the Mayo Clinic are the founding members of the   ``High Value Healthcare Collaborative (HVHC)'', which is a collective of close to 100,000
physicians and close to 10 million patients across the \textcolor{black}{U.S.}
In an early   project, HVHC found strikingly different costs and processes for total knee replacements among four hospital sites, with one site performing significantly  better than the others~\cite{harvard2014}.  Subsequently, this site’s best practices were shared with the other three and  all four could reduce their lengths of stay for knee-replacement procedures by a full day~\cite{tomek2012collaborative}.

The University of Michigan has announced a large Big Data Science Initiative targeting
health issues  in the context of mobility and wearable devices~\cite{Lampe2015}.

The University of Washington Tacoma has developed the ``RiskO-Meter''  using data analytics.
It provides a
risk score to clinicians and patients to predict the return of congestive heart failure patients to the hospital within the critical 30 day readmissions window~\cite{Naegele2015}.

{Penn Medicine}, part of the University of Pennsylvania Health System,  is working on a Big Data project to develop predictive analytics to diagnose deadly illnesses. The backbone is a homegrown enterprise data warehouse, called \textit{Penn Data Store}. An example is the prediction of \textcolor{black}{the} danger  of  {severe sepsis}, which  relies on \textcolor{black}{an} analysis of
six vital sign measurements and lab values.  The  model takes into
account more than 200 clinical variables and enables  Penn Medicine to detect 80 percent of severe sepsis
cases  \textcolor{black}{as much as 30 hours before onset of septic shock (as opposed to just two hours prior, using traditional identification methods)}~\cite{Penn2016}.

%
%\textsc{PUT SOMEWHERE ELSE !!!!!}
%
%Deep Learning is one of the most exciting developments in machine learning in recent years. It is a field that attracts amazing talents
%with stunning successes in a number of applications.
%One of the driving forces is DeepMind, a London based company owned by Google.
%DeepMind Health  is a project in which UK NHS medical data is analyzed.
%The agreement gives DeepMind access to  healthcare data on more than a  million patients~\cite{dmh2016}.
%A first outcome is the mobile app Streams, which presents timely information that helps nurses and doctors detect cases of acute kidney injury.
%% By comparing  patients’ information with millions of other cases,  the app might be able to predict that they are in the early stages of a disease that has not yet become symptomatic. Additional test can then be run to determine
%% if the prediction is correct.
%
%Two other  notable Deep Learning efforts are Deep Genomics\footnote{http://www.deepgenomics.com/} and Entlitic \footnote{http://www.enlitic.com/}. Considering the overwhelming success of Deep Learning in image analysis, one can expect that many companies will follow. One of the first approaches for using Deep Learning  (more specifically recurrent neural networks) to model the sequential decision processes in clinics is described in~\cite{esteban2015}.
%
%\textsc{AtomNet: A Deep Convolutional Neural Network for Bioactivity Prediction in Structure-based Drug Discovery}

\subsection{Comments on the Value of Big Data Studies}
\label{sec:confound}

Often the goal of Big Data studies is to draw causal conclusions, e.g., on the effectiveness of a drug  or on a possible disease cause,  and one needs to consider   the value of an observational   Big Data study versus classical randomized controlled trials (RCT).

Prospective RCTs are often cited as the gold standard for evidence since by a careful study design, effects of hidden confounders can be minimized.
But RTCs also have  their shortcomings, in particular \textcolor{black}{} due to the way patients are selected for a study and due to the small sample size.
 RCTs are often done in relatively healthy homogeneous groups of patients, \textcolor{black}{which are
 healthy except for the condition of interest, free of common diseases like  diabetes or high blood pressure,  and are  neither extremely young or old~\cite{Bresnick2015-3}.}
 If patients have several problems, treating them as if they \textcolor{black}{were mutually independent might be bad in general, and information on treatment-treatment interactions are} not be easily assessable  through RCTs. Also, interplay between diseases like hypertension, high cholesterol and depression might not become apparent in RCTs.
Since patients are difficult to recruit in general \textcolor{black}{} and the management of clinical studies is costly, sample size is often small. For the same reasons, findings need to be general and not personalized and  there are long delays until a result is certain and can become clinical practice.
 It has been suggested that patient-reported outcome measures are often  better predictors of long term prognosis~\cite{Forum2015}.
 Non-randomized,  quasi-experimental studies  are sometimes employed but provide less evidence than RCTs~\cite{harris2006use}.

%\textsc{NEW
%In contrast, some delegates challenged society’s reliance on large-scale clinical trials and saw this as a barrier to progress. Professor Knowles, for example, highlighted the considerable investment in phase IV (post-marketing authorisation) trials and postulated that this was due to the inappropriate use of RCTs in phase III. Participants also discussed the need to engage with regulators on the issue of accepting qualitative information on patient outcome as an acceptable output. It was suggested that patient reported outcome measures are better predictors of long term prognosis than many of the measures currently used and will become more prevalent as citizens contribute to data collection. Several delegates proposed a move towards alternative methodologies, such as observational studies, and data, such as qualitative information, to test the efficacy of decision rules and treatments.}

Big Data analyses, in contrast, consider data from a large variety of patients and potentially can draw conclusions from a much larger sample.
\textcolor{black}{Data} are  based on the natural population of patients, and  conclusions can be personalized.
For instance, with depressed diabetic patients, one would want to compare hospitalization rates between those taking antidepressants and those who were not,  to determine if more patients should receive psychiatric treatment to help them manage their health. Currently such studies involve great efforts.
In a  future Big Data healthcare  these questions could be answered by a simple database query~\cite{Rubenfirel2013}.

   % In contrast, a Big Data analysis will mostly concern observational studies (cohort studies, case-control studies)

Big Data analysis  mostly concerns observational \textcolor{black}{studies whose} conclusions  are considered by some to be statistically less reliable.
The main reason is that hidden  confounders might produce correlations, independent of a causal effect.  Confounders are variables that both influence {clinical decisions}  and, at the same time, outcome.
\textcolor{black}{One solution to minimize the effects of confounders  are multivariate models  where
predictive models  contain all those variables as inputs that were used in the  decision making by the caregiver.}
Unfortunately, some of these variables might not be available for analysis, such as patient symptoms and patient complains, which both are often not well documented.

   % Similarly, to analyse $\textit{clinical decisions}$ confounders are all factors that influence both the disease cause and the disease.
%As a simple example for a hidden confounder, consider that  the head physician always picks the most promising cases and the junior physician gets the difficult cases. If both physicians have a tendency to prescribe different treatments, then there is a bias that the treatment prescribed by the head physician appears more successful, simply because the head physician  got the better patients!

% \textsc{These are causal questions. Also: quasi experiments, instrumental variable methods.}

Data collection  might introduce various forms of biases.
Examples are  batch effects, which might occur  in the  merging of data from different institutions; batch effects can be addressed   by a careful statistical analysis~\cite{huang2016path,fan2014challenges}.

It is still unclear  if physicians are ready to use  evidence from Big Data.
Generally accepted is the generation of novel hypotheses by Big Data studies,  which are then clinically validated, although clinicians are critical towards  hypothesis fishing~\cite{dahl2008data}.
Of course clinical studies are very expensive and would only be initiated with significant evidence from data and with  the prospect of large  benefits.

A desired and well accepted  outcome is the discovery of novel patient subgroups  based on risk of disease, or response to therapy, using diagnostic tests.  \textcolor{black}{These subgroups  can then be the basis for a targeted therapy in  precision medicine (see Section~\ref{Personal}).}
For example,  asthma is largely regarded as a single disease and  current treatment options tend to address its symptoms rather than its underlying cause. It is now accepted  that  asthma patients can be grouped according to patterns of differential gene expression and clinical phenotype with group specific therapies~\cite{Forum2015}.

A predictive or prescriptive analysis might output a prediction (e.g., prediction of some clinical end point),  or a
ranking or prioritization of treatments.
 \textcolor{black}{Here} the  output might have been  calculated based  on many patient dimensions  and this process might be difficult  to interpret.
Prioritization is currently still  contrary to medical tradition and it remains to be seen if the medical profession will accept \textcolor{black}{this aspect} of a Big Data decision support system.

It is important to understand why machine learning solutions typically work with many inputs. In a perfect situation  a diagnostic test can reveal the cause of a problem and the subsequent therapies solve \textcolor{black}{or at least alleviate} the problem.
In reality,  even with all advances in diagnostics,  we are often still very far from being able to completely describe the health status of an individual.
\textcolor{black}{Technically, the health status of a patient consists of may aspects and only some of these  (i.e., some infections, some cancer types) can be inferred by specific diagnostic tests.}
In Big Data analysis one is partially doing ``new medicine'', i.e., one might address problems from \textcolor{black}{novel} disease subgroups or syndromes that cannot be detected unambiguously with existing diagnostic tests.
Since the statistical model then implicitly needs to infer the latent causes from observed \textcolor{black}{surrogates,} the models often become high-dimensional, and their  predictions become difficult to interpret by humans, although  predictive performance might be excellent.
This is an effect observed in  a multitude   of predictive machine learning applications in and outside of healthcare.
% Consequently, this can  mean that   clear hypotheses might not be formed, which is contrary to healthcare tradition.
% Latent causes might be the reasons why a medication sometimes seems to work and sometimes not.
The Big Data perspective is: \textcolor{black}{ If there are latent diseases, disease subgroups or  syndromes,  they leave traces in  a large number of  observable dimensions.} 

  \section{Out with the Data}
\label{External}

\subsection{Introduction}

In this section we focus on data that \textcolor{black}{are} leaving the clinic systems, i.e., data that \textcolor{black}{are} accessible to the payers, data that  \textcolor{black}{are} collected in registries and data that \textcolor{black}{are} reported to healthcare agencies.
 Payers have a unique longitudinal view on patients and can perform statistical analysis on treatment efficiencies and outcome \textcolor{black}{---}  for the optimization of their offerings, but also   for the  detection of fraud.
 Registries are  valuable sources for epidemiological research.
We will discuss Health information exchange (HIE), which  refers to various activities around  the mobilization of healthcare information electronically across organizations~\cite{wikiHIE}.
  Data reported to healthcare agencies can be used for quality control and for  policy optimization.
 As an example of the latter, we discuss the HITECH act, which is an attempt to improve the clinical system in the U.S. by encouraging an adoption of the EHR and its meaningful use via incentive programs.
Finally, we discuss privacy and data safety.

\subsection{Data Accessible to Payers: Billing Data}

The most common situation where  data \textcolor{black}{are}  leaving the clinic is when claims are filed with a payer, e.g., a health insurer or a health plan.
Depending on the particular reimbursement rules in place, payers see data of varying levels of detail,  quality and  biases.  Unfortunately,  claims data may not fully reflect a patient’s burden of illness~\cite{tang2007comparison,rosenman2013agreement}.
While the appropriateness of billing data to clinical research is often debated, many, many studies have used these data to guide clinical care, policy and reimbursement.

Claims data \textcolor{black}{deliver a holistic view of patients} across providers for  a specific  period of time, and it permits a patient centric view on health.
Claims data also \textcolor{black}{deliver} direct and indirect evidence of outcome, e.g., by analyzing readmissions,  and inform on  cost efficiencies  and treatment quality across providers.
Payer organizations are increasingly interested in better understanding their customers, in this case their patients.
Surveys, questionnaires, call center data, and increasingly social media including tweets and blogs  are analyzed for gaining insights to improve quality of services and to optimize offerings.

A major concern is the detection and prevention of abuse and fraud.
Healthcare fraud in the U.S. alone involves tens of billions of dollars of damages each year~\cite{chandola2015fraud} and fighting fraud is one of the obvious activities to immediately reducing healthcare costs.
Note that some forms of fraud actually do not only harm the payer but directly the patient (e.g., by unnecessary surgery)~\cite{Sullivan2009,betz2012experiences}.
 Naturally there is a grey zone between charging for justified claims on the one side and abuse and fraud on the other side. Certainly, billing for services never provided, e.g., for fictitious  patients or deceased patients, is clearly fraud, but if an expensive treatment is necessary or not in a case  might  be debatable.
\textcolor{black}{A 2011 McKinsey report stated that fighting healthcare fraud with Big Data analytics  can be quite effective~\cite{manyika2011big}.}

Technical solutions focus on the detection of known fraud patterns, the prioritization of suspicious cases  and the identification of new forms of fraud.
\textcolor{black}{More sophisticated  approaches use}   statistical models of clinical pathways and best practices to detect abnormal claims (against the population) and analyzes  suspicious temporal changes in charging patterns within the same provider.
In addition  one can analyze different kinds of provider networks, where nodes are the providers and the links are common patients, analyzing homophily or ``guilt by association'' patterns.
   Another measure is the black listing of providers.

  Most commercial systems use a combination of different strategies~\cite{chandola2015fraud}.
 Despite all these efforts, and mostly due to the fragmentation in the system and a huge grey zone, it is estimated that only a few percentage of the fraud actually occurring is currently being detected.

\subsection{Registries}

Disease or patient registries are collections of secondary data related to patients with a specific diagnosis, condition, or procedure.

There exist registries for  dozens of problems\footnote{http://www.nih.gov/health/clinicaltrials/registries.htm}; the best known ones are cancer registries, which have become an invaluable tool for understanding and detecting cancer within the U.S. but also \textcolor{black}{in} many other countries.

{Population-based cancer registries  regularly  monitor the frequency of new cancer cases (so called incident cases) in well-defined populations.
The basis are case reports  collected  from different sources, e.g., treatment facilities, clinicians and pathologists, and death certificates.}
If an unexpected increase of cases can be observed in registries, hypotheses about possible causes are generated.
These  are then investigated in a second step by collecting more detailed data and performing further analysis.
Registry data  \textcolor{black}{are critical to determining  geographic and temporal cancer clusters and they} can be used
for the development and tracking of the most effective therapies and  treatments.
Population based registries can also monitor the effects of preventive measures. Public health officials  use the data to make decisions on research funding and educational and screening programs~\cite{chapman2013}.

{In contrast to population registries, hospital  registries are  traditional means
 for research within a clinic or a clinic system using more detailed data about diagnosis, therapy and outcome.}

 % { As in the more recent  Big Data projects discussed in the last section, the data is used
%  for the development of  decision support systems, to determine  optimal treatments, and  for planning therapies.}

The quality of the conclusions that  can be drawn from cancer registries  critically depends on the completeness and the quality of data.
Both might  improve through the adoption of the EHR: Stage~2 of the HITECH act calls  EHR reporting to cancer registries to support  comparative effectiveness research.
In October 2012, the University of Kentucky launched a  first U.S. working model for EHR reporting of cancer cases to a state’s cancer registry~\cite{chapman2013}.

An important aspect is to guarantee that the electronic data transfer is safe and that proper precautions and safeguards have been implemented.
If only summaries are reported, HIPAA violations (see Section~\ref{sec:hipaa}) can be avoided.
Note that with registries one obtains in-use data and one needs to be aware of possible confounders distorting the analysis (see discussion in Subsection~\ref{sec:confound}).

\subsection{HIE}
\label{sec:hie}

{Health information exchange (HIE) refers to various activities around  the mobilization of healthcare information in digital form and across organizations~\cite{wikiHIE}.
It is intended to regulate the electronic transfer of clinical and  administrative information across diverse and often competing healthcare organizations~\cite{dixon2010framework}.
HIE is also useful \textcolor{black}{for} public health authorities to assist in analyses of the health of the population}

{Several organizations have emerged to support the  health information exchange efforts, both on independent and governmental/regional levels. These organizations develop and manage a set of contractual conventions and terms and develop and maintain HIE standards.}

{There are two main models for HIE data architectures. In a \textit{centralized HIE} there is a central (or master) database which holds a complete copy of all of the records of every involved patient.
In a \textit{federated HIE} each healthcare provider is responsible for maintaining the records of their individual patients, \textcolor{black}{as well as for}  data availability and common data standards.

Patient consent can be managed by an \textit{opt-in} approach or an \textit{opt-out} approach.
In opt-in,  a patient is not automatically enrolled into the HIE by default and generally must submit written permission for their data to be exchanged.

In opt-out,  patients give implicit consent  when they agree to use the services of a healthcare provider who is submitting data into an HIE. In this latter model \textcolor{black}{a  patient} can request to opt-out of the HIE, generally with a written form.

A major goal is a nationwide health information network that will allow physicians quick access to their patients' complete  medical histories without compromising their privacy. \textcolor{black}{Another benefit is  that } the data can be used  to support the learning healthcare system~\cite{lhcs2012,ONC2015}.

\subsection{Care.Data}

{The U.K. has a national health service (NHS), which attempts to address many of the problems associated with the fragmented systems in the U.S. and in many other countries.
A program called care.data was announced by the Health and Social Care Information Centre (HSCIC) in Spring 2013.  The care.data program was advertised to integrate health and social care information from different sources to  analyze  benefits and potential shortcomings of the NHS~\cite{wikiHSC}.
The data could be  used in anonymized form by healthcare researchers, managers and planners, but also by  parties from outside the
 NHS such as academic institutions or commercial organizations.}

Stated goals of the project were
\begin{itemize}
   \item to better understand diseases \textcolor{black}{} and treatments,
   \item to understand patterns and trends in public health and disease to ensure better quality care,
   \item to plan services that make the best of limited NHS budgets,
      \item to monitor the safety of drugs and treatments,
      \item and to compare the quality of care providers in different areas of the country.
\end{itemize}

Regardless of the question if the program was managed well or not,  the experience shows which type of acceptance problems projects like these can encounter.
An opt-out model was implemented where individuals  were being informed that data on their health may be uploaded to HSCIC unless they objected,  but the opt-out option was unclear.
It was seen as a major problem that it  was impossible for a patient to determine what the data will be used for, i.e., it was impossible to limit the use only for medical research by excluding insurance companies and pharmaceutical industry.}
Another issue was that the data was pseudomyzed, i.e., a unique patient identifier was used, and critics argued that this would not  be   a major hurdle for re-identification. People were worried that data \textcolor{black}{were} made accessible to consulting companies like McKinsey or PWC as well as pharmaceutical companies, like  AstraZeneca. There was also concern   that the police could access the data.

{In October 2014 the program was reviewed by the Cabinet Office Major Projects Authority and it was concluded
 that it had ``major issues with project definition, schedule, budget, quality and/or benefits delivery, which at this stage do not appear to be manageable or resolvable''.}

\subsection{Incentive Programs}

The wording is dramatic:  Some argue that healthcare is undergoing the most significant changes in its history, driven by the spiraling cost of care, shifting reimbursement models,  and changing expectations of the consumer.
Reforming the  healthcare  system to reduce the rate at which costs have been increasing  while sustaining its current quality might be critical to many industrialized countries.
An aging  population and the emergence of new, more expensive treatments will accelerate this trend.

It has been argued that by far the greatest savings could be achieved by population wide healthier lifestyles, which would largely prevent cardiovascular diseases  and  chronic conditions like diabetes.
Chronic conditions account for an astounding 75\% of healthcare costs in the U.S.~\cite{oecd2015,mckethan2009improving}. There is some hope that the proliferation of fitness and health apps might be greatly beneficial to population health (see Section~\ref{Continuous}).

Population health management tries to improve the situation by different measures such as   a  value-based reimbursement system causing providers to change the way they bill for care. The goal is to align incentives with quality and value.
Instead of providers being paid by the number of visits and tests they order (fee-for-service), their payments are increasingly based on the value of care they deliver (value-based care).
For those providers and healthcare systems that cannot achieve the required scores, the financial penalties and lower reimbursements will create a significant financial burden.

An important instrument in the U.S. is the Health Information Technology for Economic and Clinical Health Act, abbreviated HITECH Act.
It was enacted under the American Recovery and Reinvestment Act (ARRA) of 2009. Under the HITECH Act, the United States Department of Health and Human Services (HHS) is spending several tens of billions of U.S.-dollars to promote and expand the adoption of health information technology to enable a nationwide network of electronic health records (EHRs).
This \textcolor{black}{network} can then be the basis for informed population health management and for improving healthcare quality, safety, and efficiency, in general.

The general goals are to improve care coordination, reduce healthcare disparities, engage patients and their families, and improve population and public health, by, at the same time,  ensuring adequate privacy and security.

The implementation is in three stages.   An organization must prove to have successfully implemented and used a stage for a minimum of time before being able to move to a higher stage. If stages are successfully reached,  financial incentives in Medicaid and Medicare are being paid. If stages are not reached, financial penalties can be implemented by  both systems.

In Stage~1, the participating institutions  do not only need to introduce an EHR but also need to demonstrate their meaningful use.
The core set of requirements include the use computerized order entry for medication orders, the implementation of  drug-drug, and drug-allergy checks, and the implementation of  one clinical decision support rule.
Also the protection of the  electronic health information (privacy \& security) needs to be demonstrated.

Stage~2 introduces new requirements, such as demonstrating the ability to electronically exchange key clinical information between providers of care and patient-authorized entities.
 Health information exchange (HIE) (see Subsection~\ref{sec:hie}) has emerged as a core capability for hospitals for Stage~2.

Stage~3 of meaningful use is shaping up to be the most challenging and detailed level yet for healthcare providers. Among the elements are additional quality reporting, clinical decision support and security risk analysis.
The Stage~3 rule lists clinical decision support as one of the eight key objectives. Unlike the Stage~1 which required one clinical decision support rule, Stages~2 and 3 specifically require the use of five clinical decision support interventions.

Although welcomed by many, there also has been criticism of HITECH related to the increased reporting burden and the focus on reporting requirements and not on outcomes.

The HITECH act provides many opportunities for \textcolor{black}{analytics}, for example in the development of certified tools which  \textcolor{black}{provide} evidence that a provider is fulfilling the various meaningful use criteria.

Other incentive programs have been put in place as well.
For example the Centers for Medicare \& Medicaid Services (CMS) provide incentives via the Hospital Readmission Reduction Program (HRRP).
Incentives are paid if patients are not admitted to the same clinic within 30 days of release for the same problem.

The New York State Department of Health has instituted the Delivery System Reform Incentive Payment Program with the goal of transforming NY Medicaid healthcare delivery to reduce avoidable hospitalizations by 25\%.\footnote{DSRIP: http://www.health.ny.gov/health\_care/medicaid/redesign/docs/dsrip \_project\_toolkit.pdf)}
More than \$8 billion will be paid out in incentive and infrastructure payments to 25 Preferred Provider Systems (PPSs) provided they meet this ambitious goal in 5 years. The 25 PPSs are each geographically local networks of varying size (from 100+ to near 500+) including hospitals, physician practices, imaging centers, \textcolor{black}{} rehab, and hospice,  who would normally compete for patients, but have voluntarily come together to form trusted health networks (i.e., a PPS). They   have agreed to share patient data and coordinate patient care to improve patient experience through a more efficient, patient-centered, and coordinated system. The PPSs have ``signed up'' for different targeted programs (e.g., targeting mental health, fetal-maternal heath, diabetes, pediatric asthma, etc.) depending on community health assessments they performed in their area.

Although population health management might seem to be slow moving  and bland if compared to the more visible precision medicine initiatives, it has recently be argued, that the impact of  the former might be dramatically greater, if one looks at the current state of the art~\cite{bayer2015public,Bresnick2015-4}.
``Looking at
 diabetes, precision medicine may help a few scattered patients in the right clinical trials to tackle their Type~1 diabetes, but it may not prevent the 28 percent of undiagnosed Type~2 diabetics  from experiencing adverse effects from a lack of treatment the way a robust risk stratification and predictive analytics program might,'' Bayer and Galea write~\cite{bayer2015public}.

\subsection{Data Privacy, De-identification and HIPAA}
\label{sec:hipaa}

%xxxx
%
%{The Health Insurance Portability and Accountability Act (HIPAA) provides strong protections for the privacy and security of personal health information by also acknowledging that data needs to  be exchanged across organizations within a region, community or hospital system~\cite{vest2010health,fontaine2010systematic}.}
%
%
%yyyyy

Data breaches in the medical industry happen more often than expected~\cite{Blumenthal2015}.
A wake-up call was
the  February 2015 cyber attack on Anthem Health, which affected the    personal information of  78.8 million people.
Healthcare information has considerable value in the \textcolor{black}{black} market.
Since, in general,  even a major data breach does not affect revenue,   organizations have few incentives to invest in digital security; thus, regulations are introduced to encourage security measures.

The storage, access and sharing of medical and personal information of any individual is addressed in the \textit{HIPAA Privacy Rule}.
The \textit{HIPAA Security Rule}  outlines national security standards to protect health data created, received, maintained or transmitted electronically.
The latter is  also known as ePHI (electronic protected health information)~\cite{onlinetech2015}.

The HITECH Act supports the enforcement of HIPAA requirements by introducing penalties for health organizations that violate HIPAA Privacy and Security Rules.   Any company that deals with protected health information  must ensure that all the required physical, network, and process security measures are in place and followed.

\section{The Patient in Charge}
\label{Charge}

\textcolor{black}{Patients become more active in taking charge of their own health and their health data (patient empowerment) and   leave traces that can be analyzed to better understand population health and health concerns.}
On the down side,  public  traces can also be used to the patients' disadvantage and   there is an increasing  worry about bullying and  social scoring.

\subsection{Leaving Traces}

Web-based  search is part of nearly everyone’s life and is also the preferred venue to find out about one's health issues.
Health  related research often starts with Wikipedia, which is frequently consulted on health issues by both  patients and  health professionals.
Wikipedia is undoubtedly an important source of information although quality issues have been raised~\cite{tucker2014}.
There are  a number of health specific portals (e.g., \textit{ netdoctor, healthline,  Yahoo Health, WebMD, whatnext.co} and \textit{RevolutionHealth}), some of which are managed by leading healthcare providers such as the  Mayo Clinic and the Cleveland Clinic.

Other web services help patients  find the right provider for their problems.
Among them are commercial resources like \textit{Healthgrades} and \textit{ZocDoc} as well as government resources such as Medicare’s \textit{Hospital Compare} site. One can observe an  increasing willingness  to ``shop for health''  leading to the question of which company would become the  ``Amazon of healthcare''~\cite{aoh}.

Similar to  the general population, patients are increasingly active in social networks like Facebook and various blogs.  \textcolor{black}{There are also }  a number of  social network services  addressing specific health issues~\cite{DBLP:books/crc/p/Kotov15}.
The motivation is obvious: Patients with the same problems want to  communicate and exchange information.
Problem-specific communities are organized by commercial and noncommercial web portals and special services can be provided to these groups by third parties.

Not just patients might want to organize themselves, but also clinics and healthcare professionals, and collaboration tools appear on the market.\footnote{https://bps-healthcare.siemens.com/teamplay/, http://www.cmtcorp.com/}

%“Imaging systems generate a lot of data, however very little of it is utilised to make improvements," states Ben Reed, syngo Business Manager, GB & Ireland at Siemens Healthineers. “The teamplay network is intended to help hospitals make the most of this data, analyse it and exchange it with other experts, forming the basis for prompt and well informed decision-making. This can be in a clinical, financial or operational sense."

\subsection{Analyzing Traces}

Statistics on anonymized search query logs and traces in social media  can be analyzed to inform public health, epidemiologists  and policy makers.
\textit{Infodemiology} is a new term standing for the large-scale  analyses of anonymized traces, which  can potentially yield valuable results and insights.
\textcolor{black}{Infodemiology  can support
the  early detection of epidemics, the analysis and modeling  of the  flow of illness
and other purposes~\cite{horvitz2015data}.
It can  address public health challenges and can} provide new avenues for scientific discovery~\cite{horvitz2015data}.

A widely discussed example is the analysis of  search query logs as indicators for disease outbreaks.
The idea  is that   social media and search logs might indicate an  outbreak of an infectious disease like a flu immediately, including detailed temporal-spatial information of its spread.
Previously, such outbreaks  might go unnoticed for days of even weeks.
But models have proven difficult. \textcolor{black}{Google Flu Trends} for example, predicted well initially but the fit was very poor later~\cite{lazer2014parable,olson2013reassessing}.

Another application is the  detection of adverse drug reactions, which  could be improved by jointly analyzing data from the U.S. Food and Drug Administration’s Adverse Event Reporting System, anonymized search logs and social media data~\cite{horvitz2015data}.
The analysis of patients' traces has increasing importance in  \textit{pharmacovigilance}, which  concerns the  collection, detection, assessment, monitoring, and prevention of adverse effects with pharmaceutical products.
Still there is little experience yet in the quality, reliability and biases in  data generated from Web query logs and social network sites and conclusions should be drawn with great caution~\cite{janssens2012research,kalf2013predictive}.

There is also a \textcolor{black}{danger in patients leaving traces in social media:}
When re-identified,  traces can be aimed at making inferences about unique individuals that could be used to infer   \textcolor{black}{their health status}.
Many problems are associated, e.g.,  with  social scoring in healthcare.
\cite{horvitz2015data} reports on a Twitter suicide prevention application called Good Samaritan that monitored individuals’ tweets for words and phrases indicating a potential mental health crisis.
The service was removed after increasing complaints about violations of privacy and imminent dangers of stalking and bullying.
As pointed out by~\cite{horvitz2015data}, health issues can also be inferred from seemingly unrelated traces.
Simply changing communication patterns on social networks and internet search might  indicate a new mother at risk for postpartum depression.

{Another issue is that  some   companies are working together with analytic experts to track
 employees’ search queries, medical
claims, prescriptions and even voting habits to get insight into their personal lives~\cite{Wilkinson2016}.
Although  HIPAA legislation forbids  employers  to
view their employees' health information, this  does not  apply to third parties.
A company which received public attention  is
 Castlight, which  gathers data on  workers’ medical information, such as  who is  considering pregnancy or who may need back surgery.\footnote{http://www.castlighthealth.com/} Castlight's policy is to only inform and advice the individuals directly and only report statistics to employers. }

\textcolor{black}{Patient privacy issues} are increasingly addressed by regulators, e.g., in the \textcolor{black}{U.S.} by the \textit{Americans with Disabilities Act} (ADA) and  the \textit{Genetic Information Non-Discrimination Act} (GINA).

\cite{horvitz2015data} points out the technical difficulties in protecting the citizens against  violations, in the face of powerful machine learning algorithms which can ``jump categories'':  Machine learning can enable  inferences about health conditions  from nonmedical data generated far outside the medical context~\cite{horvitz2015data}.

\subsection{PatientsLikeMe}

An openly commercial  social network initiative  is  PatientsLikeMe~\cite{gupta2011patientslikeme,brownstein2009power} with several hundred thousands of  patients using  the platform and addressing more than a thousand  diseases.
The majority of users have neurological diseases such as ALS, multiple sclerosis and Parkinson’s, but PatientsLikeMe  is also increasingly addressing   AIDS and mood disorders~\cite{frost2008social,wicks2010sharing}.

PatientsLikeMe is not merely a chat board with self-help news but also collects quantitative data.
It has designed several  detailed questionnaires which are   circulated regularly to its members.
For example, epileptics can enter their seizure information into a seizure monitor.
It has a survey tool to measure how closely patients adhere to their treatment regimen, but also scans language in the chat boards for alarming words and expressions.
\textcolor{black}{ PatientsLikeMe offers a number of  services.
Together with the Massachusetts Eye and Ear Hospital it created a contrast sensitivity test  for people with Parkinson’s and  hallucinations. }% that come with mood disorders.
 %  The stated goal is the realization of a learning healthcare system.

The business model of PatientsLikeMe is not based on advertising.
Instead, the company has based its business model \textcolor{black}{on} aligning patient interests with industry interests, i.e.,  accelerated clinical research, improved treatments and better patient care.
To achieve these goals, PatientsLikeMe sells aggregated, de-identified data to its partners, including pharmaceutical companies and medical device makers.
In this way,  PatientsLikeMe aims to help partners in the healthcare industry better understand the real-world experiences of patients as well as the real-world course of disease.
Some of PatientsLikeMe’s past and present partners include UCB, Novartis, Sanofi, Avanir Pharmaceuticals and Acorda Therapeutics.

\subsection{Managing Your Own Data}

Consumers might not only want to research their health issues and communicate with others,
 but also possibly  manage their own data.

If patients take responsibility for their own data, they must be able to store, manage and control the access to \textcolor{black}{these} data.
By nature, this would overwhelm \textcolor{black}{the patients’} capabilities and commercial and noncommercial services realize some of \textcolor{black}{the} necessary  functions~\cite{sunyaev2010evaluation}.
The core is a personal health record (PHR) which is a patient centered assembly of all personal health information.

Among the earliest offerings is the  Microsoft HealthVault, which  addresses \textcolor{black}{individuals} who want to manage their own or their family's health.
The HealthVault permits the storage and consolidation of a patient’s life health information and \textcolor{black}{enables the patient} to give access to this information to selected parties.
For example, the HealthVault   keeps digital records of children's immunization records or an \textcolor{black}{individual’s}  medical imaging results and displays them to authorized parties whenever wanted. Doctors can send data and files right into an individual’s HealthVault account.
The site lets the users generate  letters that can be sent to their healthcare professionals, outlining instructions\textcolor{black}{, } and security and encryption details.
As discussed in the next section, a lot of healthcare and fitness related data are produced by mobile devices,   and services like HealthVault offer convenient functionalities for managing, storing  and analyzing those data. The World Medical Card and WebMD \textcolor{black}{offer} related services.

Naturally, due to privacy issues and their distributed nature, PHRs are difficult to use as part of an analytics project; nevertheless the rich information in a PHR can be used in a personalized advisory and alarming system. 
 \section{Continuous Healthcare}
\label{Continuous}

With the tremendous  technological progress and prevalence of   mobile devices, the disruptive potential of mobile health \textcolor{black}{(mHealth)},  \textcolor{black}{and also,   more general, } technology-enabled care,  is frequently \textcolor{black}{being} discussed~\cite{collins2012fulfill,kay2011mhealth}.
 A new generation of affordable sensors is able to collect health data outside of the clinic in an unprecedented quality and quantity.
  This enables the transition from \emph{episodic} healthcare, dominated by occasional encounters with healthcare providers, to \emph{continuous} healthcare, i.e., health monitoring and
 \textcolor{black}{care,   potentially} anytime and anywhere!
 Continuous healthcare certainly has the potential to create a shift in the current care continuum from a treatment-based healthcare to a more prevention-based \textcolor{black}{system: Many} health problems can be prevented by a healthy life style and the early detection of disease onset, in combination with early intervention.
 However, the full potential remains to be unlocked as a 2012 Pew Research Center study about mobile health \textcolor{black}{reveals~\cite{Fox2016}: While} about half of smartphone owners use their phone to look up health information, only 1 in 5 smartphone users own a health app.
 Currently this exciting field is in a flux and opportunities, challenges and crucial factors for \textcolor{black}{its} widespread adoption are discussed in current research ~\cite{gagnon2016m,free2013effectiveness,aranda2014systematic,miyamoto2016tracking}.

\subsection{Technological Basis}

The technological basis of mHealth includes smart sensors, smart apps and devices, advanced telemedicine networks such as the optimized care network\footnote{http://www.optimizedcare.net/} and supporting software platforms.
There is a broad range of new devices that have entered the market: smartphones, smart watches, smart wrist bands, smart head sets and Google Glass, among others.  In the future, patient-consumers might  use a number of different devices that measure a multitude of different signals: "\textcolor{black}{headsets} that measure brain activity, chest bands for cardiac monitoring, motion sensors for seniors living alone, remote glucose monitors for diabetes patients, and smart diapers to detect urinary tract infections"~\cite{Blumenthal2015}.

\textcolor{black}{A Body Area Networks (BAN) is}   another form of \textcolor{black}{a} technological enabler with sensors that measure physiological signals, physical activities, or environmental parameters and \textcolor{black}{it comes} along with an internet-like infrastructure.
 BANs \textcolor{black}{are, for example, being} used to monitor cardiac patients and help to diagnose cardiac arrhythmias~\cite{jovanov2011body}.

 Add-ons to mobile devices such as lab-on-a-chip technologies are particular interesting technologies and might represent a new form of point-of-care devices. ~\cite{laksanasopin2015smartphone} presents a laboratory-quality immunoassay that can be run on a smartphone accessory and~\cite{knowlton2015sickle} present a 3D printed attachment for a smartphone for the detection of sickle cell disease.

Especially for developing countries with a limited infrastructure the potential of such technologies is tremendous.

From an engineering perspective, continuous healthcare  is related  to    condition monitoring and predictive maintenance, enabled by smart sensors, connectivity and analytics --- a combination often referred to as the internet of things (IOT).
By measuring and aggregating the signals \textcolor{black}{from} many different persons, machine learning algorithms can be trained  to \textcolor{black}{detect,  e.g., } anomalies and unexpected correlations that \textcolor{black}{might} generate new insights.
Open source initiatives such as the Open mHealth\footnote{http://www.openmhealth.org/} initiative are important enablers that could pave the way to overcome the data \textcolor{black}{integration challenges.  }

\subsection {Use Case Types}

\subsubsection{Disease prevention} \textcolor{black}{Smartphones are increasingly being used for measuring, managing and displaying lifestyle and health  parameters,   related, e.g., to  weight, physical activity, smoking, and diabetes. }
Improving lifestyle and fitness of the general population has the potential to reduce healthcare costs dramatically and thus  \textcolor{black}{fitness and health  monitoring} might have dramatic positive impact on \textcolor{black}{both} population health and healthcare cost.
In a recent statement, the American Heart Association (AHA) reviews the current use of mobile technologies to reduce \textcolor{black}{cardiovascular disease} (CVD) risk behavior~\cite{burke2015current}.
CVD continues to be the leading cause of death, disability and high healthcare costs~\cite{burke2015current} and is thus a prime example for investigating the potential of mHealth technologies.
The work investigates different tools available to consumers to prevent \textcolor{black}{CVD,} ranging from text messages (e.g. smoking cessation support), wearable sensors, and other smartphone applications.
While more evidence and studies are needed, it appears  that mHealth in CVD prevention is promising. The AHA strongly encourages more research.

\subsubsection{Early detection}
Many diseases can be treated best when discovered early and before they cause serious health consequences.
Early detection can happen at the population level or at the individual level. ~\cite{collins2012fulfill} highlights an early warning system for disease outbreaks caused by illness related parameters such as environmental exposure or infectious agents.
On the individual level, the previously mentioned Body Area Network is a major enabler for early detection of abnormalities.
So-called smart alarms can be understood as another form of early detection on the individual level.

\textcolor{black}{Smart alarms cover a range of applications, such as the  monitoring of  heart activity, breathing, and potential falls,
 and are especially relevant to  the elderly}~\cite{jovanov2011body}.

The company AliveCor\footnote{http://www.alivecor.com/} is offering a mobile ECG that is attached \textcolor{black}{to} a mobile device (either smartphone or tablet).
The attached device creates an ECG that is then recorded via an app.
The \textcolor{black}{mobile} ECG is cleared by the FDA and can also detect atrial fibrillation,  a leading cause of mortality and morbidity.
AliveCor states that  the device has been used to record over five million  \textcolor{black}{ECGs, and that  these data are then} the basis for training an anomaly detection algorithm.

\subsubsection{Disease management} Healthcare costs can be reduced when the patient \textcolor{black}{is} monitored at home instead of in the clinic and if physicians can optimize care without the need to call in the patients for a medical visit.
Some hospitals and clinics collect continuous data on various health parameters as part of research studies~\cite{Blumenthal2015}.
Especially the management of chronic diseases can benefit from continuous healthcare. In a recent review~\cite{hamine2015impact}, the authors screen systematically for randomized clinical trials that give evidence about better treatment adherence when using mHealth technologies.
\textcolor{black}{Applications} range from simple SMS services to video messaging with smartphones and other wireless devices. They conclude that there is,  without doubt,  high potential for these technologies but,  as the evidence in the trials was mixed, further research is needed  to improve  usability, feasibility, and acceptability.

\subsubsection{Support of translational research} With hundreds of millions of smartphones in use around the world, the way patients are recruited to participate in clinical studies might change dramatically.
 In the future patients might be able to decide themselves if they want to participate in a medical  study and they might be able to  specify how their data will be used and shared with others.

Major research institutions have already developed apps for studies involving asthma, breast cancer, cardiovascular disease, diabetes and Parkinson's disease.
One interesting use case  is the control of disease endpoints in clinical trials with mHealth technologies.
As a concrete example,  Roche developed an app to control or measure the clinical endpoints of Parkinson disease.\footnote{http://www.roche.com/media/store/roche\_stories/roche-stories-2015-08-10.htm}
The app, which complements the traditional physician-led assessment, is currently used in a Phase I trial to measure in a continuous way disease and symptom severity.
The app is based on the Unified Parkinson’s Disease Rating Scale \textcolor{black}{(UPDRS), which} is the traditional measurement for the disease and symptom severity.
The test, which takes about 30 seconds, investigates six \textcolor{black}{endpoint-relevant} parameters \textcolor{black}{and involves a} voice test, balance test, gain test, dexterity test, rest tremor tests and postural tremor.

The Clinical Trials Transformation Initiative\footnote{http://www.ctti-clinicaltrials.org/}, an association representing diverse stakeholders along the clinical trial space,  \textcolor{black}{is envisioning the }  next generation of clinical trials.
Recently, the initiative has launched a mobile clinical trials program to investigate how mobile technologies and other off-site remote technologies can further facilitate clinical trials.

\subsection{Selected Projects}

Many different projects \textcolor{black}{have  begun} involving clinics, research institutes and technology providers.
In a recently started pilot between the MD Anderson Cancer Center and Polaris Health, Apple \textcolor{black}{Watches are collecting} data from breast cancer patients~\cite{Crotti2015}.
According to a Polaris statement, data to be collected \textcolor{black}{include treatment} side effects, information about sleep behavior, levels of physical activity, and patient \textcolor{black}{mood.}
Researchers will combine this information with EHR data from the patients and health data of other breast cancer patients to create new insights.

Another example \textcolor{black}{demonstrates the great potential} for developing countries.
\textcolor{black}{Medic Mobile, \footnote{https://medicmobile.org/} a non-profit technology company,} has developed a software platform that is used in 23 countries in Africa, Latin America, and Asia to improve care in rural areas.
The use cases of the platform range from antenatal care, childhood immunization, disease surveillance, and drug stock monitoring.
For antenatal care,  the organization reports on their homepage that approximately 500.000 people have been covered in the countries Bangladesh, Kenya, and Nepal.
Over 1.800 community health workers are using their smartphones to register women in a central database once they are pregnant. Automated text messages are \textcolor{black}{then sent to organize appointments, } and health workers can register any potential danger signs.

Japan Post, one of Japan's largest insurers, joins forces with IBM and Apple to address issues of an aging generation~\cite{Herper2015}.
They will be designing app analytics and cloud services around the smartphone to help to connect millions of seniors with their families but also to healthcare services.
The project will help \textcolor{black}{Japan Post to both know} more about its customers and to improve the health and wellness of its seniors, thus allowing \textcolor{black}{individuals} to live potentially longer, healthier and more independently.

% Communities have been established that even try to put their lives into data.
The Quantified Self movement~\cite{swan2013quantified} uses sensors to put a person's daily life into data by self-tracking biological, physical, behavioral or environmental signals~\cite{swan2013quantified}. The community is supported by a company of the same name.\footnote{http://quantifiedself.com/}

\subsection{Implications for the Clinical Setting and the Doctor's Office}

{Some hospitals and insurers have already recognized the willingness of patients to use telemedicine services~\cite{Heather2015} \textcolor{black}{and  are} offering video consultations ---a contemporary ``house call''--- to patients via Skype and other internet conferencing systems.
 ``In the way that video calls and instant messaging revolutionized the way people communicate with others, now health systems are exploring how e-health consultations for routine ailments can relieve the pressure on primary care systems that are functioning beyond capacity,'' Blumenthal writes~\cite{Blumenthal2015}.
 Some patients find these e-visits to be cheaper and more convenient \textcolor{black}{than visits to their doctor's office.}
  About 55\% of patients recently asked in a  survey would send a photo of their skin to a dermatologist for consultation.\footnote{http://www.pwc.com/us/en/health-industries/healthcare-new-entrants/assets/pwc-hri-new-entrant-chart-pack-v3.pdf}
 Researchers say more evidence \textcolor{black}{is} needed to understand if virtual medical visits will actually reduce costs or improve health outcomes. But the demand among patient-consumers is there and some large insurers have begun to pay for these virtual consultations~\cite{Blumenthal2015}.}

\subsection{Regulatory Implications}

% The recent growth and increased popularity of mHealth applications in combination with questions of regulatory oversight of these has created a sphere of action where many different stakeholders have to align with each other.

 The continuous healthcare ecosystem has brought together stakeholders that were previously more or less unconnected and now have to interact.
 For instance in the \textcolor{black}{U.S.}, certain app developers suddenly  have to deal with premarket notification or so-called 510(k) clearance processes from the FDA.
 The driving question here is which type of mHealth applications fall under FDA's jurisdiction over medical devices.
 Indeed,  different classifications of ``mobile medical applications'' and according FDA \textcolor{black}{guidances} now exist, but they do not appear to be  \textcolor{black}{finalized yet.}

While it is the traditional responsibility of the
FDA   to oversee the safety and effectiveness of medical devices (\textcolor{black}{also including} certain types of mobile apps),   some politicians and industry representatives are afraid that innovation is hampered by regulatory oversight.
However, first warning letters to doctor's had to be sent out where mobile medical apps showed unexpected behavior; another case revealed that about 52 adverse event reports \textcolor{black}{in the FDA’s reporting database} were generated for one specific diabetes app within two years~\cite{hamel2014fda}.
Clearly,  further intensive dialogue between stakeholders is needed.~\cite{hamel2014fda} \textcolor{black}{describes} in detail the challenges that come along with the regulation of \textcolor{black}{mHealth technologies,  together with} potential alternative regulatory scenarios.

%Some mHealth products are beginning to appear
%in the FDA’s reporting database.43 A search
%of this database for adverse events related to just
%one diabetes app (iBGStar) returned 52 reports
%from 2012 through early 2014.

\subsection{Conclusion}

In conclusion, the potential benefits of continuous healthcare are tremendous. Of course many challenges remain: will it be possible to solve major issues with data privacy?
Will it be possible to maintain population interest in the long run or are health apps just a temporary phenomenon?
Will there be \textcolor{black}{sustainable} business models? Will it be possible to find \textcolor{black}{technical solutions for  the data integration challenges?  What will be the eventual clinical impact of
 digital mHealth?}

 \section{Getting Personal}
\label{Personal}

\subsection{Precision Medicine is Changing Healthcare}

Maximizing the positive effect of a healthcare intervention by  concurrently   minimizing adverse side effects has always been the dream of individualized healthcare.
Over the last decades it became clear that this goal cannot be achieved with insights from conventional studies alone, which have been focusing on empirical intervention efficacy and  side effects in large patient study groups.
The reason is that, due to the biological diversity of individuals, environment and pathogenesis, any incident of a complex disease is like no other.
Precision medicine, personalized medicine, individualized medicine and stratified medicine ---terms we will use interchangeably---   all refer to the grouping of patients based on risk of disease, or response to therapy, using diagnostic tests.
Precision medicine \textcolor{black}{thus} refers to the idea to customize healthcare,  with medical decisions, practices, and procedures being tailored to a  patient group.
In its most extreme interpretation,
this leads to the ``n=1'' principle, meaning that therapy should be tailored to  the patient’s individual characteristics, sometimes referred to as the ``unique disease principle''~\cite{ogino2013molecular}.

%In this article we will use the term precision medicine interchangeably with personalized medicine, individualized medicine and stratified medicine.

% Many times molecular patient or disease characteristics are the basis of a patient or patient group characterization.

 Without question, the most important milestone for the realization of a personalized medicine
 was  the publication of the reference sequence of the human genome
 about 15 years ago~\cite{lander2001initial, venter2001sequence}.
 In the following years, the patient’s  genomic profile,  supplemented with  other molecular and cellular data,  became the basis for a  dramatic progress in the understanding of the molecular basis of disease.
 The impact of this knowledge is not limited to research: As new  analytical methods like next generation sequencing  (NGS) and  new proteomic platforms bring cost down,  molecular data will increasingly  become part of clinical practice.

 \textcolor{black}{With growing data sets,
 increasingly complex phenomena, even with weak  associations, can be discovered and validated. Genome-wide association studies (GWAS)\textcolor{black}{,} with more than a million attributes collected from
up to several thousands individuals\textcolor{black}{,} are good examples.
The main goal is to  link  {the generated} data to clinically  {actionable} information.
  The vision of a \textit{real-time personalized  healthcare} is the
rapid and real-time analysis of biomaterials obtained from the patients based on newest research results in a network of research labs and clinics.
Research and clinical  applications  go along with a huge increase in  volume and variety of data available to characterize the physiology and pathophysiology.}

The insights in the biological causes of disease might lead to a more meaningful categorization  of disease, at some point in the future replacing medical  codes, which were mostly developed based on clinical phenotyping~\cite{Forum2015}.

%\textsc{Better understanding; redoing the way diseases are classified (ICD phenotypes to true molecular causes); data driven categories;  	Taxonomy based on phenotypes !!!!!}

By far the greatest efforts in precision medicine have been devoted to cancer (oncology), but precision medicine becomes increasingly relevant to other medical domains,
 e.g.,  the  central nervous system (e.g., Alzheimer's and depression),  immunology/transplant,    pre-natal medicine, pediatrics, asthma, infectious diseases and cardiovascular~\cite{kulkarni2013personalized}.

\subsection{Understanding Disease on a Molecular Level}

\textcolor{black}{In the last decades, a lot of attention has been focusing on understanding the genetic causes of disease.}

\textit{Monogenetic} disorders with a high penetrance have been linked to mutations \textcolor{black}{of single inherited genes.} The causative genes of most monogenic genetic disorders have now been identified~\cite{duncan2014revolution}.

Monogenetic diseases are relatively rare and attention has shifted largely to \textit{complex diseases}: Most common diseases, including most forms of cancer, are based on an interaction of several  factors including a number of inherited genetic variations, one or several mutations acquired during cell life time, as well environmental factors.
Consider, for example, that worldwide approximately 18\% of cancers  are related to infectious diseases~\cite{anand2008cancer}.
Due to the complex interplay of several factors, these diseases show, what has been termed,  ``missing heritability''.

% GERMLINE
Insights into inherited genetic  cell disorders are  obtained from germline DNA, typically obtained from blood cells.
Genome wide association studies (GWAS) examine the correlation between germline genetic variations and common phenotypic characteristics, such as breast cancer~\cite{visscher2012five}.
\textcolor{black}{The {likelihood of a person developing a disease} in their lifetime can sometimes be predicted according to germline DNA profiles, permitting early intervention and possibly preventing an outbreak of the disease.}
\textcolor{black}{With the establishment of next generation sequencing (NGS), \textcolor{black}{in the future}  the whole genome  might be
decoded  for costs} in the order of a few hundred U.S. dollars \textcolor{black}{and this will make clinical genome analysis much more common. }
\textcolor{black}{Eventually, the increasing use of genome sequencing will lead to better insights into which diseases can be explained by genetic variance and could revolutionize molecular medicine for some diseases.}

Additional genetic variations of interest are those acquired during the lifetime of  somatic cells,
which comprise all cells that form an organism's body, excluding the germ cells.
As genetic alterations accumulate, the somatic cell can turn into a malignant cell and form a cancerous tumor.
  Genetic profiles (mutations and amplifications) of somatic cancer cells are obtained from analyses of tumor biopsies.
   \textcolor{black}{Distinct mutations} and gene amplification \textcolor{black}{patterns can be  linked to} clinically relevant characteristics, such as prognosis or therapy response~\cite{cancer2012comprehensive}.
    In some cases the tumor is easily accessible, however in other cases,  \textcolor{black}{as for tumors or metastases of certain organs (e.g. brain, liver, lung),}  a biopsy is not standard of care. In those cancer patients,  the access to the material from which the genomic information could be obtained is difficult.
  Recently, novel methods have been developed  that permit the analysis of  alternative sources of tumor material, such as  circulating tumor cells (CTCs). These are cancer cells that  have shed into the blood stream from a primary tumor.
  CTCs can constitute seeds for subsequent growth of additional tumors (metastasis) in distant organs, triggering a mechanism that is responsible for the vast majority of cancer-related deaths.
  \textcolor{black}{The analysis of CTCs has been called  a ``liquid biopsy''.
   Also circulating tumor DNA (ctDNA) was found to resemble the tumors genomic profile, being useful for cancer detection and prediction of therapy efficacy~\cite{bettegowda2014detection}.}

% RNA
So far we have been focusing on \textcolor{black}{DNA. The} transcription of RNA from DNA is called gene expression. This step plays a crucial functional role, because RNA is translated directly into functional proteins. Furthermore RNA has regulatory functions, of which many are not yet \textcolor{black}{fully} understood.
 Transcriptomics is the study of transcriptomes (RNA molecules, including mRNA, miRNA, rRNA, tRNA, and other non-coding RNA), their structures and functions. DNA microarrays (which, despite their name, really test for RNA) and  RNA-seq (RNA sequencing) can reveal a snapshot of RNA presence and quantify cellular activities at a given moment in time.
 In some cancers,  such as breast cancer,  the expression of \textcolor{black}{some} genes has already been proven to be of great clinical \textcolor{black}{relevance.
 Increasingly,  genomewide} gene expression analyses are becoming available to characterize cancer diseases~\cite{nevins2007mining}.

 % From RNA levels one can conclude which proteins are currently being generated in a cell, however not, whether these proteins have functional role in the cell, the organ or the disease.}

% Proteins
Whereas the genome contains the code, the proteins are the body's functional worker molecules.
Several methods like \textcolor{black}{immunohistochemistry and} enzyme-linked immunosorbent assays (ELISA) are used in clinical practice \textcolor{black}{for protein analysis}.\footnote{This is a test that uses antibodies and color change to identify a substance.} In research\textcolor{black}{,} and recently also in clinical tests, mass spectroscopy is used to determine many proteins in a tissue, opening this field for high throughput and big data approaches~\cite{catenacci2015mass}.  \textcolor{black}{Increasingly, protein} microarrays are used  as a high-throughput method  to track the interactions and activities of many proteins at a time.

% Other OMICS
While the transformation of genetic information into functional proteins is recognized as being clinically highly relevant, the clinical relevance of other ``omics'' fields is still under investigation.
 Epigenomics, metabolomics and lipidomics are three further levels of systems biology which might be unraveled by big data analyses. Epigenetic changes modify genes on a molecular level, \textcolor{black}{such that expression is altered; the analysis of the effects of these modifications  is part of current research.}
 Metabolomics  concerns chemical fingerprints that specific cellular processes leave behind, in particular, the study of their small-molecule metabolite.
 Lipidomics focuses on cellular lipids, including the modifications made to a particular set of lipids, produced by an organism or system.

The  environment is increasing the number of possible interactions that play a role in the etiology (i.e., disease cause) and pathogenesis of a disease\footnote{http://www.genome.gov/27541319}.
 The exposome encompasses the totality of human environmental (i.e. non-genetic) exposures from conception onwards, complementing the genome.
\textcolor{black} { Disease often involves several factors. }
 For example, scientists believe that, for most people, Alzheimer's disease results from a combination of genetic, lifestyle and environmental factors that affect the brain over time.\footnote{http://www.mayoclinic.org/diseases-conditions/alzheimers-disease/basics/causes/con-20023871}
Only in less than 5 percent of cases, Alzheimer's is caused by specific genetic changes that, by themselves,  virtually guarantee a person will develop the disease.\footnote{Reality is even more complex: there is \textcolor{black}{also}  heterogeneity within a particular tumor. The hypothesized \textit{cancer stem cell model} asserts that within a population of tumor cells, there is only a small subset of cells that are tumourigenic (able to form tumours). These cells are termed cancer stem cells (CSCs), and are marked by \textcolor{black}{their} ability to both self-renew \textcolor{black}{and differentiate. One assumes} a process of natural selection within a given tumor which also would explain why cancer is so difficult to fight:  a treatment  might eliminate one strain giving room for another strain to develop. It has been argued that this could be a major problem for the vision on a personalized medicine~\cite{Tannock2016}.
 An alternative but related explanation is the \textit{clonal evolution model}. }

As a medical field,  molecular medicine is concerned with the molecular and genetic problems that lead to diseases and with the development of molecular interventions to correct them.
A better understanding of the underlying molecular mechanisms of diseases can lead to great advances in diagnostics and therapy.
In particular, cancer subgroups can be determined by omics profiles and the most effective treatment with smallest adverse effects can be determined for each subgroup.
This concept is at the center of precision medicine.

%Tumour heterogeneity describes the observation that different tumour cells can show distinct morphological and phenotypic profiles, including cellular morphology, gene expression, metabolism, motility, proliferation, and metastatic potential.[1] This phenomenon occurs both between tumours (inter-tumour heterogeneity) and within tumours (intra-tumour heterogeneity).
%
%There are two models used to explain the heterogeneity of tumour cells. These are the cancer stem cell model and the clonal evolution model. Intratumour heterogeneity (ITH) is a prevalent feature in multiple cancer types and constitutes a considerable challenge to the optimization of anticancer therapies. Tumour heterogeneity describes the observation that different tumour cells can show distinct morphological and phenotypic profiles, including cellular morphology, gene expression, metabolism, motility, proliferation, and metastatic potential.

% Breast cancer:
{To give insight in what is clinically relevant \textcolor{black}{today,  consider}  the {concrete example of \textcolor{black}{breast} cancer}.
Molecular techniques \textcolor{black}{have  changed} our understanding of the basic biology of breast cancer and provide the foundation for new methods of ``personalized'' prognostic and predictive testing.
Several molecular markers are already established in clinical practice such as high penetrance breast cancer causing genes (\textit{BRCA1} and \textit{BRCA2})~\cite{miki1994strong,wooster1995identification}.
Also the characterization of the tumor is driven by molecular markers such as estrogen receptor, progesterone receptor and a genetic alteration, the \textit{HER2} amplification~\cite{slamon1987human}.
% All of these characteristics are well known from times before high throughput molecular analysis or the achievement of genomewide data approaches
 Since the  biological signals of those markers are quite strong, they were discovered already in the 90's of the last century, even before high throughput molecular analysis  became a reality.
 Now, more than 15 years after the primary publication of the human genome,  \textcolor{black}{many  levels} of biology (DNA, RNA, Protein, Epigenetics, miRNA, \ldots) can be analyzed at relatively low cost\textcolor{black}{,  revealing  detailed} and comprehensive insight into the biology of a cell\textcolor{black}{.}

 A particular role for {understanding} breast cancer {on the molecular level} \textcolor{black}{play the efforts around ``The Cancer Genome Atlas'' (TCGA).
   It was one of the first Big Data efforts that}  compared the genetic information of the tumor with the genetic information of the blood on a large scale for each single of the three billion base pairs.
   {See also Section \ref{bigDinR}}. This project could, for the first time,  describe systematically, which genes will mutate in the course of the pathogenesis of a healthy mammary cell to a breast cancer cell~\cite{cancer2012comprehensive}. }

\subsection{Molecular Diagnostics and {Drug} Therapy}

% The correlation with clinical meaningful data such as treatment response or disease classification   is as important as the collection of molecular data of patients.

%\textcolor{black}{The alignment of clinical and molecular data in  integrative data systems  to improve  disease understanding and patient treatment is one }   of the next great challenges.

\textcolor{black}{The need for a precision medicine is {quite} apparent  when looking at the limited drug response rates,  as published research from the early 2000s  reveals \cite{spear2001clinical}}. Thus  alternatives to  the traditional ``blockbuster'' \textcolor{black}{models are needed}~\cite{Forum2015}.

 %Allen Roses, of GlaxoSmithKline, who is quoted  as saying more than 90\% of drugs only work in 30-50\% of people.\footnote{http://news.bbc.co.uk/2/hi/health/3299945.stm}.} when looking at drug  response rates from the early 2000s as published research revealed \cite{spear2001clinical}.}
 %Thus \textcolor{black}{\sout{the age of the  ``blockbuster'' drug might be over since it no longer %provides an appropriate model for how the pharmaceutical industry will operate in the future} alternative operating models than the traditional ``blockbuster'' model are currently % discussed}~\cite{Forum2015}.

 % At the same time, hospital beds are occupied with patients that exhibited  adverse drug reactions.  }

%\textsc{pharmacogenetics}
%
%\textsc{Severe Adverse Drug Reaction; drug resistance}
%
%\textsc{Only 50\% of patients benefit from a spec drug: need biomarkers!!!
%}
%
%
%
%•	90 \% of gsk drugs only work in 30-50\% of patents
%•	8000 beds with ADR (adverse drug reactions)!
%
%\textsc{Professor Holgate reiterated the notion that the ‘blockbuster’ drug no longer provides an appropriate model for how the pharmaceutical industry will operate in the future.}

%\textsc{Enabling clinicians to select targeted therapies, often by acting as ‘companion diagnostics’ to particular stratified medicines.}

% VOLKER Biomarker
The \emph{diagnostic} part of precision medicine heavily relies on biomarkers.
In molecular diagnostics, the term  biomarker refers to any of a patient's molecules that can be measured to assess  health and  that can be obtained from  blood, body fluids, or tissue.
\textcolor{black}{Biomarker testing is at the center of personalized medicine and tests are specific, e.g., to DNA, RNA or protein variations.
 Biomarkers may   test if certain proteins \textcolor{black}{are} overactive, in particular if they help to promote cancer growth  and  therapy, and  may be based on  the identification of a molecule (a \textit{drug target}), often a protein,  whose activity needs to be  modified by a drug.}

 {Pharmaceutical research tries to find drugs, so called \textit{targeted drugs},  that bind the \textcolor{black}{drug target} with the goal to influence underlying disease mechanisms}.
% Targeted drugs are typically  small-molecule drugs (ligands).
% Volker: Targeted Therapy
\textcolor{black}
{\emph{Targeted therapy} uses  a number of  different strategies {to fight  tumors}.
Some targeted drugs block (inhibit) proteins that are signals for cancer cells to grow. Drugs called angiogenesis inhibitors stop tumors from making new blood vessels, which greatly limits \textcolor{black}{their growth}.
Immunotherapy is a  treatment that uses the body's own immune system to help fight cancer, e.g., uses the patient's immune system to attack tumor  cells.
A strategy is to generate   {antibodies}  (e.g.,  monoclonal antibodies)
  which are man-made versions of large immune system proteins that bind to very specific target proteins on cancer cell membranes.
To give an example,  the protein HER2   is a member of the human epidermal growth factor receptor  family and its overexpression plays an important role in certain forms of breast \textcolor{black}{cancer; HER2} is the target of the monoclonal antibody trastuzumab.}

 \textcolor{black}{Biomarkers are  relevant in  companion diagnostics, which  are diagnostic tests  used as  companions} to a therapeutic drug to determine its applicability, e.g., efficacy and safety,  to a specific patient.\footnote{http://www.fda.gov/MedicalDevices/ProductsandMedicalProcedures/ InVitroDiagnostics/ucm407297.htm}
  Companion diagnostics are \textcolor{black}{co-developed} with drugs to aid in selecting or excluding patient groups for treatment.
 %   with that particular drug on the basis of their biological characteristics that determine responders and non-responders to the therapy.

% repurposing
{While most drugs have been approved for very specific  diseases, they might also sometimes be effective in other diseases.  One  reason is that  the targets in both diseases  might have  the same alterations.
% especially in cancer  it has become obvious that targeted therapies might work as well in another form cancer.
% This leads to the question whether a medication can be used in other diseases or cancers
% if the target has the same alteration.}
  % The introduction of a  specific drug for a disease  is simplified if the drug had previously  been admitted for treatment for  another  disease.
  The application of known drugs and compounds to treat new indications  is called drug repurposing. Analytics can play a role in finding good candidates~\cite{rastegar2015toward}, \cite{aliper2016deep}.
A well known case is the pain medicine Aspirin, which was found to be effective in treating and preventing heart disease. {In cancer, as another example, it could be shown that a drug that works against a mutated gene in melanoma is also active in other cancers if the respective mutation in BRAF is found \cite{hyman2015vemurafenib}.
} The main  advantage of drug repositioning over traditional drug development is that  \textcolor{black}{---since the repositioned drug has already passed a significant number of toxicity and other tests---} the drug's  safety is known and the risk of failure for reasons of adverse toxicology is reduced. Thus, the introduction of a  specific drug for a \textcolor{black}{new} disease  is greatly  simplified.

\subsection{Implementing Precision Medicine}

% Difficulties in Reality

{As a major milestone, a first insurer has begun to cover the cost of the sequencing of the full  germline and tumor genomes of cancer patients\footnote{http://www.reuters.com/article/ca-nanthealth-idUSnBw116104a+100+BSW20160111}.
Despite \textcolor{black}{the great perspectives of precision medicine,} it still faces many challenges}.
The implementation  will require changes and improvements on many levels,  reaching from technology developments (one genome can comprise up to 400GB of data) over social and ethical challenges to legal implications and the need for large scale educational programs for patients, physicians, researchers, healthcare providers, insurance companies and even politicians~\cite{green2011charting}.

The abundance of data and possibilities \textcolor{black}{to join information sources} raises the question, whether current  rules for intellectual property, reimbursement and personal privacy have to be adapted to   personalized medicine.

Regulatory authorities have already acknowledged those challenges and released a report: ``Paving the Way for Personalized Medicine: FDA’s role in a New Era of Medical Product Development'' ~\cite{fackler2009paving}. In this report the FDA describes a framework of  how to integrate genomic medicine into clinical practice and drug development. \textcolor{black}{Steps to implement precision medicine include the development of regulatory scientific standards, research methods,
and reference material}~\cite{fackler2009paving}.
\textcolor{black}{Implementing and  commercializing} precision medicine will demand new standards with regard to the  protection of \textcolor{black}{patients' privacy and that of their families.}
\textcolor{black}{Data protection issues} arise especially for healthy individuals who have genetic predisposition for a disease or patients who have a genetic alteration (either germline or somatic) and who are thought to be non-responsive to standard treatments:  In some cases the person, for which the molecular data \textcolor{black}{were created, might not want to know the complete  interpretation of those results.}
An important milestone regarding privacy issues in the \textcolor{black}{U.S.} was the Genetic Information Nondiscrimination Act (GINA) in 2008 that protects American citizens from being discriminated based on their genetic information with respect to employment and health insurance.

\subsection{Big Data in Molecular Research}\label{bigDinR}

% HUGE DATA IS GENERATED CORRELATION WITH CLINICAL DATA

% BEGIN FIX !!!!!!!!!!!!!!!!!!!!!!!!!!!!!!!!!!!!!!!

% RESEARCH GWAS
The aim is to use the newly gained insight into etiology, pathogenesis and progression of diseases for novel treatments and prevention.
Large international consortia were formed over the last years integrating data from
not  seldom \textcolor{black}{several} hundreds of thousands of individuals to compare genetic and environmental information of healthy individuals with diseased patients.
Several of those consortia have built super-consortia merging data and biomaterials of several large scale \textcolor{black}{consortia.}
One example is the OncoArray Network\footnote{http://epi.grants.cancer.gov/oncoarray/} GWAS study, in which more than 400,000 individuals are genotyped for more than 570,000 genetic variants. Diseases included in this effort are breast cancer, ovarian cancer, colon cancer, lung cancer and prostate cancer.
 \textcolor{black}{GWAS studies  examine the correlation between germline gene variations and phenotypic characteristics and  explain a certain amount of attributable risk for a disease within a population.}
For the case of  breast cancer, GWAS led to the discovery of around 100  risk genes~\cite{michailidou2013large}.
For the individual the statistical effects are rather small and implementation into healthcare is highly dependent on programmes which would utilize this information in an epidemiological way, i.e. \textcolor{black}{by} selecting patients for individualized prevention or early detection of a disease.
 This \textcolor{black}{strategy}  requires tens if not hundreds of thousands or millions individual decisions \textcolor{black}{in a population}, which will require \textcolor{black}{highly scalable Big Data technology.}

The 1000 Genomes Project~\cite{10002010map}, launched in  2008, was an effort  to sequence the genomes of at least one thousand anonymous participants.
Many rare variations  were identified, and eight structural-variation classes were analyzed. It is followed by the  100,000 Genomes project, which was launched in  2013. It aims to sequence 100,000 genomes from UK's NHS patients by 2017 and it is  focusing on patients with rare diseases \textcolor{black}{and more} common cancers.\footnote{https://www.theguardian.com/politics/2013/jul/05/health-jeremy-hunt}.
{An interesting and less costly alternative is the  distributed collection of genomic data from patients who donate their decentrally analyzed genome to central projects.\footnote{http://datascience.columbia.edu/donate-your-genome-science-learn-more-about-your-ancestry-health}$^{,}$\footnote {http://www.personalgenomes.org/}$^{,}$\footnote{https://dna.land/}. From a data management perspective,  these decentralized approaches require  innovative ways of storing and analyzing huge amounts of data employing  distributed computing\footnote{https://arvados.org/}.}

Biobanks  are great sources for molecular research. Biobanks store biological samples (often cancerous tissue) for use in research  like genomics and personalized medicine~\cite{hirtzlin2003empirical}.

%The 100,000 Genomes Project is a UK Government project that is sequencing whole genomes from National Health Service patients. The project is focusing on rare diseases, some common types of cancer, and infectious diseases.\footnote{https://www.theguardian.com/politics/2013/jul/05/health-jeremy-hunt}
%

% This will be another challenge in times of clinical big data application.

%
%% RESEARCH GWAS
%One of the problems is that, to achieve this analysis in an empirical approach, { huge sample sizes} have to be achieved, which is a challenge to achieve a high level of data quality.
%
%As mentioned above in the field of molecular epidemiology large consortia and super-consortia have been build including several hundreds of  thousands of individuals. As for breast cancer risk factors this effort led to the discovery of around 100 more risk genes~\cite{michailidou2013large}.

%\textcolor{black}{\sout{Unconventional projects, such as the decentralized collection of genomic data from patients who donate their de-centrally analyzed genome to central projects might be an approach  to save resources \textcolor{black}{\footnote{http://datascience.columbia.edu/donate-your-genome-science-learn-more-about-your-ancestry-health}$^{,}$\footnote {http://www.personalgenomes.org/}$^{,}$\footnote{https://dna.land/}}. From a data management perspective,  these decentralized approaches require  innovative ways of storing and analyzing huge amounts of data with distributed computing\footnote{https://arvados.org/}.}}

% RESEARCH GWAS
As stated before, complex diseases involve a number of causes. Unfortunately, to study the interaction of disease causes involving, for example, several gene variations requires even larger sample sizes.  Similarly, the study of complex patterns behind the spatio-temporal \textcolor{black}{disease progression}  requires the acquisition and management of huge data samples~\cite{soon2016abstract}.

%
%{Combining risk factors and investigating interactions for biological effects or to combine genetic alterations with gene expression or protein function will need {even higher sample sizes} and approaches, which still have to be developed for the collection of clinical and molecular data.  While this in itself is a great challenge there might be even more
% RESEARCH GWAS
% correlation with clinical data
%The correlation with clinical meaningful data such as treatment response or disease classification   is as important as the collection of molecular data of patients. The alignment of clinical and molecular data in integrative data systems  and improvements using this data for disease understanding and patient treatment will be among  of the next great challenges.

% END FIX !!!!!!!!!!!!!!!!!!!!!!!!!!!!!!!!!!!!!!!!!!!!!!!!!!

\subsection{Digitization Challenges in Precision Medicine}

Recent publications~\cite{Hayden2015,stephens2015big} estimate that storage needs for molecular data  will exceed by far those of Twitter or YouTube, which is of great concern to researchers  and healthcare professionals alike.

This perception is supported by the many large scale population-based initiatives (e.g. the aforementioned Genomics England 100K project or the NIH precision medicine initiative) that will collect genomic and other biomedical data from individuals \textcolor{black}{for} the next 5-10 years.
A comprehensive and recent overview of these cohort studies from publicly or private funded entities can be found in \cite{huang2016path}.
The experiences gained from these initiatives will reveal interesting insights and lessons learned about data management of genomic and other ``omics'' data (e.g. transcrpitomics, proteomics, metabolomoics, epigenomics), emerging standards, and data privacy topics such as informed consent.

To \textcolor{black}{consistently improve} patient outcome and medical value, it will become very important to bridge the gap between all the previously mentioned ``omics''  data and clinical outcome.
Indeed \textcolor{black}{clinical sequencing}  for advanced patient diagnosis is becoming more and more common, but many questions still \textcolor{black}{remain,} e.g.,  \textcolor{black}{what parts of the genomic data should become part of the EHR records?}
Here, important consortia such as \textit{emerge} (Electronic Medical Records and Genomics) and \textit{CSER} (Clinical Sequencing Exploratory Research) will hopefully pave the way towards a more integrated view of genomics in the clinic \cite{shirts2015cser}.
Structuring, organizing, synchronizing different terminologies across clinical data repositories is the prerequisite to make clinical data meaningful.
In that context companies such as Flatiron Health have developed powerful tools and processes to \textcolor{black}{tackle  data integration challenges} and offer structured knowledge bases that can yield new insights.\footnote{http://fortune.com/2014/07/24/can-big-data-cure-cancer/}

 % Therefore data integration concepts such as ontologies will become crucial to organize the clinical data mess.

\textcolor{black}{In many current efforts,  data are aggregated  across many patients with the goal of  developing   Clinical Decision Support (CDS) systems.}
The American Society of Clinical Oncology (ASCO) launched a program named CancerLinQ  that envisions to learn not only from trial \textcolor{black}{data  but} also from the mass of EHR \textcolor{black}{records.
A goal is that doctors get support in their decision making by matching their patients' data with outcomes of patients across the U.S.}
Patients gain \textcolor{black}{confidence if their treatment decisions are} based on their personal profile and on the shared experiences of similar cancer cases across the \textcolor{black}{U.S.}
Finally, researchers can access this massive amount of de-identified health information to generate new hypotheses for research.
To make CancerLinQ's vision happen,  several different data types and technologies have to be orchestrated ranging from longitudinal patient records, cohort analyses, quality metrics,  to interactive reporting and text analytics \cite{shah2016building}.
Interoperability between different EHR systems will be another crucial success factor for the CancerLinQ initiative.

\subsection{Traditional IT Players are Entering  Precision Medicine} % MARKUS

The outlined data management and analytics challenges in precision medicine are being addressed by  a number of established IT companies. Here are some examples.

SAP has teamed up with American Society of Clinical Oncology (ASCO) to implement CancerLinQ~\cite{abernethy2013asco,shah2016building}. SAP's in-memory technology platform SAP HANA  will play a crucial role in providing the infrastructure and algorithms to analyze the vast amounts of diverse data to provide clinical decision  support.\footnote{https://connection.asco.org/magazine/features/cancerlinq\%E2\%84\%A2-takes-big-leap-forward}

IBM with its Watson technology~\cite{ratner2015ibm,savvy2015watson} has recently started a collaboration with the New York Genome Center (NYGC) to generate and analyze the exome, complete genome data, and epigenetic data linked to clinical outcomes from participating patients. The partners plan to generate an open knowledge base using the generated data\footnote{https://www.genomeweb.com/informatics/ibm-nygc-expand-partnership-new-pilot-cancer-study}.

Dell is partnering with the Translational Genomics Research Institute (TGen) to tackle pediatric cancer in Europe and in the Middle East. In addition, Dell recently announced that its Cloud Clinical Archive \textcolor{black}{---}currently storing over 11 billion medical images and around 159 million clinical studies from multiple healthcare providers\textcolor{black}{---} \textcolor{black}{will  support storage and management of} \textcolor{black}{genomic data}. The long term goal will be to combine medical imaging diagnosis with advanced genomics to impact patient care.

Intel is also looking into the precision medicine space. Saffron,  a cognitive computing company that Intel acquired in 2015,  is studying how users can gain additional insights from above mentioned Dell's Cloud Clinical Archive. The company is also offering Natural Language Processing capabilities and the platform can be compared  to IBM Watson's offering. In addition, within the context of Barack Obama's Precision Medicine Initiative, Intel launched a Precision Medicine Acceleration \textcolor{black}{Program}\footnote{https://www.whitehouse.gov/the-press-office/2016/02/25/fact-sheet-obama-administration-announces-key-actions-accelerate}.

Microsoft also supports the \textcolor{black}{U.S.} government's Precision Medicine Initiative by hosting genomic data sets in Microsoft's Azure cloud platform by end of \textcolor{black}{2016 free of charge}. \footnote{https://www.whitehouse.gov/the-press-office/2016/02/25/fact-sheet-obama-administration-announces-key-actions-accelerate}

Amazon Web Services (AWS) is offering HIPAA-compliant cloud storage and data \textcolor{black}{security.} Therefore AWS often functions as a backbone of genomics data management platforms and several companies such as Seven Bridges or DNAnexus rely on the AWS technology. As a concrete example\textcolor{black}{,} the Cancer Genomics Cloud (CGC)\textcolor{black}{,} which includes the well-known \textcolor{black}{``The Cancer Genome Atlas''} (TCGA)\textcolor{black}{,} is operated by Seven Bridges and runs on the AWS cloud.

Alphabet Inc. is investing heavily in precision medicine.
This happens mainly either through the many investments taken by Google Ventures or by own research and development activities from subsidiaries such as Verily or Calico. Investments in companies related to precision medicine from
Google Ventures include Flatiron Health, Foundation Medicine, \textcolor{black}{and} DNAnexus among others.
\textcolor{black}{Among Google's} initiatives are,  e.g.,  Google Genomics or the Google Baseline Study.
Google Genomics is Google's HIPAA-compliant cloud platform for storing and
managing \textcolor{black}{genomic data}; besides} offering access to publicly available data sets such as \textcolor{black}{the TCGA}, customers can load their own genomic data sets and run analyses on the data through the offered API. The Google baseline study aims to collect different types of data such as molecular, imaging, clinical and data related to patient engagement to understand \textcolor{black}{patterns that are typical for  healthy individuals.}

%
%\textsc{PUT SOMEWHERE ELSE !!!!!}
%
%Deep Learning is one of the most exciting developments in machine learning in recent years. It is a field that attracts amazing talents
%with stunning successes in a number of applications.
%One of the driving forces is DeepMind, a London based company owned by Google.
%DeepMind Health  is a project in which UK NHS medical data is analyzed.
%The agreement gives DeepMind access to  healthcare data on more than a  million patients~\cite{dmh2016}.
%A first outcome is the mobile app Streams, which presents timely information that helps nurses and doctors detect cases of acute kidney injury.
%% By comparing  patients’ information with millions of other cases,  the app might be able to predict that they are in the early stages of a disease that has not yet become symptomatic. Additional test can then be run to determine
%% if the prediction is correct.
%
%Two other  notable Deep Learning efforts are Deep Genomics\footnote{http://www.deepgenomics.com/} and Entlitic \footnote{http://www.enlitic.com/}. Considering the overwhelming success of Deep Learning in image analysis, one can expect that many companies will follow. One of the first approaches for using Deep Learning (more specifically recurrent neural networks) to model the sequential decision processes in clinics is described in~\cite{esteban2015}.
%
%\textsc{AtomNet: A Deep Convolutional Neural Network for Bioactivity Prediction in Structure-based Drug Discovery}
%

All these efforts illustrate that information technology is moving quickly into personalized healthcare and therefore will be a main enabler to realize the goals of precision medicine.

% The crucial challenge is  to
% \textcolor{black}{turn  vast} amounts of collected and managed data} into knowledge and insights.

\subsection{A View to the Future:  a Truly  ``n=1''-Medicine}

% Big Data

% on what might be possible in the future: First steps towards a truly  ``n=1''-medicine by Big Data approaches} % MARKUS !!!!

% A view on what might be possible in the future: First steps towards a truly  ``n=1''-medicine by Big Data approaches

% antigen produces the immune reaction; the epitope is the important part; neoepitope: new one ot yet recoizable

% TREATING CANCER Neoepitopes

 \textcolor{black}{Dramatic improvements in the quality and speed of genomic sequencing and analysis as  clinical diagnostic tools
 for individual patients,
 combined with the innovations propelling immuno-oncology,
 are paving a new era of  truly personalizing the treatment of cancer.}
At the heart \textcolor{black}{of these prospects  are the newfound abilities} to rapidly identify and target tumor cells with specific DNA mutations unique to each cancer patient.
The products of mutated \textcolor{black}{genes,  encoding} altered proteins,  are so-called ``neoepitopes'' and serve as the molecular address to direct and redirect immune cells \textcolor{black}{for killing the tumor cells and  for procuring} long term immunity.
Neoepitopes are defined as unique genetic \textcolor{black}{alterations that} result in unique novel proteins. \textcolor{black}{They are found specifically in a patient’s tumor (but not in  normal tissue) and can be targeted by the immune system to attack the tumor with minimal off target toxicity.}

\textcolor{black}{It }is highly unlikely that the same neoepitopes occur in other patients, and if so only in small groups of patients.
\textcolor{black}{The generation of drugs  to specific neoepitopes in real-time
 is a vision of a  real-life ``n=1'' medicine~\cite{tureci2016targeting,srivastava2015neoepitopes}.}

% TREATING CANCER Neoepitopes ... Big Data
Identifying neoepitopes for each patient is made possible by high-throughput whole genome or exome sequencing and by the direct comparison of abnormal tumor DNA with each patient’s own normal DNA.
\textcolor{black} {This widens the search for drugable targets (neoepitopes) in the >99\% of the genome deemed untargetable or unimportant by panel sequencing and reduces the significantly high false positive error rates associated with tumor-only sequencing techniques~\cite{jones2015personalized}.}
\textcolor{black}{To increase the precision in individualizing treatments, which  are targeting neoepitopes,
further requires a confirmation of the expression of the  mutated genes,
thus avoiding another potential pitfall of false positive errors
and the potential for the altered protein to induce immunogenicity.}

If a tumor is found to express unique \textcolor{black}{neoepitopes,} they can serve as a ``molecular address'' for the immune system. Therefore there is a good rationale that the neoepitope can be delivered to the immune system by an immunogenic vehicle like a vaccinating virus.
One such vehicle is the adenovirus which may be engineered to express within its DNA many neoepitopes, and\textcolor{black}{,} upon injection, can locally infect dendritic \textcolor{black}{cells of the immune system} which then present an identified neoepitope to the immune effector cells and trigger an immune response against the tumor cells.
 Despite great promise, the use of adenovirus  or any other foreign delivery vehicles remains hindered due to the pre-existence or the induction of neutralizing antibodies against them by the patient’s immune system.
  This limitation has been overcome by engineered adenoviruses which are capable of safely vaccinating and re-vaccinating against hundreds of neoepitopes and tumor associated antigens despite pre-existing immunity against \textcolor{black}{adenoviruses}~\cite{morse2013novel}.
 Remarkable results have thus far been published demonstrating the delivery of \textcolor{black}{tumor-associated} \textcolor{black}{antigens by  engineered adenoviruses} in a cohort of late-stage colorectal cancer patients~\cite{balint2015extended}.

A more recent development has been the engineering and application of immune cells (T-cells and NK-cells) that express antibodies on their surface as part of a ``chimeric antigen receptor'' (CAR) for direct targeting of tumor cells expressing their cognate antigens.  One particular approach, an off-the-shelf human NK cell line dubbed NK-92, is engineerable to produce innumerable CARs. These cells are now being engineered to produce \textcolor{black}{CARs targeting neoepitopes  } discovered to be expressed by individual cancer patients’ tumor cells, thus \textcolor{black}{enabling} a novel, truly personalized immunotherapeutic approach to fight cancer.  For this and many other reasons, the discovery of neoepitopes has the potential to be a watershed moment in the war against cancer.
\textcolor{black}{These examples show that the  utilization of the immune system to fight cancer  requires yet another layer of data, \textcolor{black}{leading to a}  true ``n=1'' medicine.}

% TREATING CANCER Neoepitopes
{One of the challenges with neoepitope discovery and targeting will be the management of Big Data: teraFLOPS of compute resources in a cloud environment are required to \textcolor{black}{generate, manage and analyze} terabytes of sequencing data, including whole genome and/or whole exome sequencing, RNA sequencing and molecular modeling of immune presentation of neoepitopes.
These activities require compute and storage under HIPAA, \textcolor{black}{as well as} high-speed and large-bandwidth connectivity for rapidly transiting sequence data from sequencing labs to supercompute/cloud \textcolor{black}{environments,}  such that derivation and delivery of neoepitope targeting platforms \textcolor{black}{are} enabled in actionable time for each patient.
\textcolor{black}{Long term storage of data from multiple biopsies for each patient also needs to be provided.}
These challenges require significant infrastructure and resources,  \textcolor{black}{which are already realized by some private, Big Data} supercompute clouds interconnected by dedicated fiber infrastructure capable of transporting terabytes of data at terabits per second.
\textcolor{black}{Such infrastructures had originally  been developed} for  financial trading markets, \textcolor{black}{but are now} retrofitted to meet the needs of sequencing analysis and neoepitope discovery.}

\subsection{Outlook} % MARKUS

Realizing personalized medicine for \textcolor{black}{every patient around the world  would} result in Big Data challenges of unprecedented scale. Large investments in  computing and storage facilities are required and all stakeholders, including patients, doctors, nurses, insurers, lawmakers and the public,  need to get involved,  educated and trained.
 Privacy and safety concerns need to be addressed and  the general public needs to understand the eventual benefits of a personalized  medicine involving Big Data technologies and patient profiling.

Many efforts are underway to strengthen the role of \textcolor{black}{personalized}  medicine. Among them \textcolor{black}{is  President} Obama's
 ``Precision Medicine Initiative'' (PMI)~\cite{collins2015new}.

% There are substantial  expectations in powerful joint efforts to fight cancer are great~\cite{devita2015death}.

 \section{Assessments  and Conclusions}
\label{Conclusion}

% Wspialy exciting where completely novel data is generated: contin HC, personalized medicne

% Importance of EHR ... providing newest info about newest research at POC; data quality; statistical issues

% policy

% my  data

% continuous HC

% Pers Medicine

% GINA: discrimination !!!!!

% NEW

% \begin{itemize}
 %  \item	\textcolor{black} {Patient engagement}
 %  \item	\textcolor{black} {Accountable care}
 %  \item	\textcolor{black} {Experience / Engagement}
 %  \item	\textcolor{black} {Population Health Management}
 %  \item	\textcolor{black} {BMWI richtig?}
 %  \item \textcolor{black} {value-based care}
% \end{itemize}

It is unquestionable that healthcare will experience dramatic changes in the coming years and that   digitalization and large-scale data analytics will be among the   key technologies.
Precision healthcare \textcolor{black}{---with enormous potential for a better, more effective,  and personalized  treatment of cancer and other diseases---}  will require the  acquisition, exchange, storage and analysis of huge amounts of data generated in  research and clinical practice.
 Molecular patient data will bring new richness to patient profiles, including genome profiles and expression profiles.
The EHR has the potential to become the central
digital fingerprint of a patient
% \textcolor{black}{and will store}
% longitudinal, complete and up-to-date patient
%information.
%
%% The  EHR  will become the key document  to present valuable, complete and up-to-date patient data.
%It
 \textcolor{black}{and it will be} the basis for  optimal personalized treatment decisions.  \textcolor{black}{It will provide} valuable information for a learning healthcare system. Completeness and accuracy \textcolor{black}{of information} is a precondition that interventional causal conclusions can be derived.
A tight and timely integration of
EHR information, i.e. real-world data, will ensure that newest
findings can \textcolor{black}{immediately be} transported into clinical practice.  With readily available population data and well-defined outcome measures,  the effectiveness of a new treatment or a new drug can be evaluated rapidly and \textcolor{black}{caregivers}  can be advised to adapt accordingly.

 High volumes of  data will be  generated by continuous healthcare  which will permit the monitoring of patients with chronic problems and will generate data streams to be managed and analyzed  in real time, enabling continuous screening and early intervention.
 Trusted data centers \textcolor{black}{might} become an individual's health memory and support the management of  the health of \textcolor{black}{individuals} and their families.  They \textcolor{black}{will} enable  functionalities such as  reminders, alarming, health advice and the initiation of  preventive measures.

%
%\textsc{LEARNING FROM DATA INVOLVED CAUSAL ANALYSI !!!!}
%
%Not for control but to optiamyll hel patients!!!

% Cost Explosion
% The cost explosion in healthcare leaves no choice but to accept these developments:

\textcolor{black}{Despite  clear benefits, it is still largely unclear  how exactly  and when exactly the impact will be realized.}
  There is a lot of excitement and activity in continuous healthcare and personalized medicine but, in general, we are currently still far  from generally accepted solutions.  \textcolor{black}{Data privacy, liability and other legal concerns,  as well as viable data-driven business models,  are unsolved issues.} Despite these uncertainties, we \textcolor{black}{already} see a lot of public and private investments.

%
%
%\textsc{Continuous and molecular data most interesting since completely new data and informative in the future!}
%
%\textsc{EHR is important ... policies changes are important. }
%
%\textsc{Importance of EHR ... providing newest info about newest research at POC; data quality; statistical issues}
%
%

% \textcolor{black} {Experience / Engagement}

% User Interface
A challenging  question is  how an intelligent learning healthcare  system should interact \textcolor{black}{with the individual to achieve  engagement and provide best user experience.}
  When and how should such a system interfere with the individual's life?  Should the individual be informed on a likely positive finding?  \textcolor{black}{A commonly} discussed example is Huntington's disease for which genetic tests exist but currently no cure.
  Less dramatic but still potentially  bothersome
are health concerns  such as overweight:  How often should an individual be reminded that weight loss and exercise would increase longevity?
What is just the right level of decision support and interference in an individual's life?
Maybe be an individual does not want to \textcolor{black}{know about a condition or a problem}?
Maybe the individual does not want anybody to know?
\textcolor{black}{Patient engagement and  user
experience is increasingly getting in focus.}\footnote{https://www.dartmouth-hitchcock.org/stories/article/40037}

 % The Genetic Information Non-Discrimination Act in the U.S.  is partially addressing these issues for genetic information.

\textcolor{black}{There is also  the question in which way} a learning healthcare system should support treatment decisions.
Supplying newest research results on \textcolor{black}{a} patient's problems might  obviously be a good idea, but it is largely an open question how decision support  can be integrated into the workflow of the \textcolor{black}{caregivers}.
% Disease subgroups which were detected via analytics and confirmed by clinical studies should find acceptance, but
Can a  \textcolor{black}{caregiver} accept results from a machine learning system that uses
 high dimensional patient information but where it is difficult or impossible to explain the reasoning behind its   predictions and recommendations?

 % and whose  might not easily be explainable?
%  These issues  can make or break the success of digitalization and Big Data!

%Most people accept that treating all patients the same is bad, but maybe treating all  healthcare providers  the same is bad as well. How is it possible to  improve the average quality of care  without hurting the top performers?
%
%

% Positive forces:
In this paper we have described the state of the healthcare systems  and various  attempts to improve it \textcolor{black}{via more  effective processes,} standardized data formats, data exchange in trusted networks,  and improvements in    policies and reimbursement rules.
It is important that all involved stake holders, but in particular \textcolor{black}{caregivers} and patients,  personally experience  the benefits of the new developments and not just suffer from  the additional bureaucratic burden of,  for example,   reporting and  maintaining a high-quality  EHR: Trust must be generated and benefits must   \textcolor{black}{readily be} apparent since \textcolor{black}{the vision of a better and more efficient future healthcare}  only works with    support from all groups.

% Privacy:
Greatest concerns are clearly associated with data privacy and data security and generally acceptable solutions are not yet available.
The Genetic Information Non-Discrimination
Act in the U.S. is partially addressing \textcolor{black}{data protection} for genetic
information.
In general, one might want to  distinguish between  the privacy concerns of  patients with  severe health issues,  who might see clearer benefits from sharing their data, and  individuals
 without major health problems, who might not see immediate benefits in data sharing.
 \textcolor{black}{Privacy protection is a   very serious issue}: Imagine a hack which gives access to your complete (in the future more rich and meaningful) health record  to un-authorized parties, which  would open the door to discrimination and black mail!

% What if  foreign intelligence agencies keep a health profile of each citizen of your country or if your health information is collected by criminal organisations  to be sold to anybody who can pay the price?
%Abuse with criminal intend must be severely  punished.

% \subsection{XXX Business Models and Legal Issues}
Currently, sustainable  data-driven business models are still unclear  and new reimbursement models must be developed that are tailored towards a data-driven  medicine. The legal situation of what is allowed and what is not allowed must be clear and unambiguous, which is not the case in many countries:  \textcolor{black}{Viable business models require a solid legal basis.}

%
%building viable business models on an uncertain
%legal basis is not an option.
% Another challenging  issue is the monetary value of the data and commercial ownership~\cite{bollier2010promise}.

%How should patient consent be dealt with? How long is patient consent  valid? What solutions need FDA approval? In Europe there are a number of EU and national regulations which need to be considered and which are mutually inconsistent. These are just a few issues that need to be clarified!

\textcolor{black}{Notwithstanding the indicated challenges:} we hope
that we could convey in this paper the great potential
of digitalization and large scale data analytics for a better and
more effective patient care.

\section*{Acknowledgment}

Volker Tresp acknowledges support  by the German Federal Ministry for Economic Affairs and Energy,  technology program ``Smart Data'' (grant  01MT14001).

\bibliographystyle{IEEEtran}
%

% \bibliography{hcbd}
\bibliography{IEEE-HC-BigData}

\begin{thebibliography}{100}
\providecommand{\url}[1]{#1}
\csname url@samestyle\endcsname
\providecommand{\newblock}{\relax}
\providecommand{\bibinfo}[2]{#2}
\providecommand{\BIBentrySTDinterwordspacing}{\spaceskip=0pt\relax}
\providecommand{\BIBentryALTinterwordstretchfactor}{4}
\providecommand{\BIBentryALTinterwordspacing}{\spaceskip=\fontdimen2\font plus
\BIBentryALTinterwordstretchfactor\fontdimen3\font minus
  \fontdimen4\font\relax}
\providecommand{\BIBforeignlanguage}[2]{{%
\expandafter\ifx\csname l@#1\endcsname\relax
\typeout{** WARNING: IEEEtran.bst: No hyphenation pattern has been}%
\typeout{** loaded for the language `#1'. Using the pattern for}%
\typeout{** the default language instead.}%
\else
\language=\csname l@#1\endcsname
\fi
#2}}
\providecommand{\BIBdecl}{\relax}
\BIBdecl

\bibitem{biesdorf2014healthcare}
S.~Biesdorf and F.~Niedermann, ``Healthcare’s digital future,''
  \emph{McKinsey \& Company}, 2014.

\bibitem{manyika2011big}
J.~Manyika, M.~Chui, B.~Brown, J.~Bughin, R.~Dobbs, C.~Roxburgh, and A.~H.
  Byers, ``Big data: The next frontier for innovation, competition, and
  productivity,'' \emph{McKinsey Global Institute}, 2011.

\bibitem{conger2012data}
K.~Conger, ``Data deluge: mastering medicine’s tidal wave, chapter {B!g}
  data. what it means for our health and the future of medical research,''
  2012.

\bibitem{kayyali2013big}
B.~Kayyali, D.~Knott, and S.~Van~Kuiken, ``The big-data revolution in {US}
  health care: Accelerating value and innovation,'' \emph{Mc Kinsey \&
  Company}, 2013.

\bibitem{wikiHIE}
\BIBentryALTinterwordspacing
{Wikipedia}, ``Health information exchange --- {W}ikipedia{,} the free
  encyclopedia,'' 2016. [Online]. Available:
  \url{https://en.wikipedia.org/wiki/Health-information-exchange}
\BIBentrySTDinterwordspacing

\bibitem{lhcs2012}
\BIBentryALTinterwordspacing
{National Academies of Sciences{,} Engineering{,} and Medicine}, ``The learning
  health care system in {A}merica,'' 2012. [Online]. Available:
  \url{http://www.nationalacademies.org/hmd/Activities/
  Quality/LearningHealthCare.aspx}
\BIBentrySTDinterwordspacing

\bibitem{rind2011interactive}
A.~Rind, T.~D. Wang, W.~Aigner, S.~Miksch, K.~Wongsuphasawat, C.~Plaisant, and
  B.~Shneiderman, ``Interactive information visualization to explore and query
  electronic health records,'' \emph{Foundations and Trends in Human-Computer
  Interaction}, vol.~5, no.~3, pp. 207--298, 2011.

\bibitem{musen2014clinical}
M.~A. Musen, B.~Middleton, and R.~A. Greenes, ``Clinical decision-support
  systems,'' in \emph{Biomedical informatics}.\hskip 1em plus 0.5em minus
  0.4em\relax Springer, 2014, pp. 643--674.

\bibitem{bright2012effect}
T.~J. Bright, A.~Wong, R.~Dhurjati, E.~Bristow, L.~Bastian, R.~R. Coeytaux,
  G.~Samsa, V.~Hasselblad, J.~W. Williams, M.~D. Musty \emph{et~al.}, ``Effect
  of clinical decision-support systems: a systematic review,'' \emph{Annals of
  internal medicine}, vol. 157, no.~1, pp. 29--43, 2012.

\bibitem{Bresnick2015-2}
J.~Bresnick, ``Healthcare big data analytics: From description to
  prescription,'' \emph{Healthcare IT Analytics}, 2015.

\bibitem{Blumenthal2015}
S.~Blumenthal and G.~Somashekar, ``Advancing health with information technology
  in the 21st century,'' \emph{Huffpost Healthy Living}, 2015.

\bibitem{SAS2015}
{SAS}, ``Applying data to improve patient-centric and personalized medicine,''
  \emph{SAS white paper}, 2015.

\bibitem{DBLP:books/crc/p/Hersh15}
\BIBentryALTinterwordspacing
W.~R. Hersh, ``Information retrieval for healthcare,'' in \emph{Healthcare Data
  Analytics.}, 2015, pp. 467--505. [Online]. Available:
  \url{http://www.crcnetbase.com/doi/abs/10.1201/b18588-17}
\BIBentrySTDinterwordspacing

\bibitem{DBLP:books/crc/p/RahmanR15}
\BIBentryALTinterwordspacing
R.~Rahman and C.~K. Reddy, ``Electronic health records: {A} survey,'' in
  \emph{Healthcare Data Analytics.}, 2015, pp. 21--59. [Online]. Available:
  \url{http://www.crcnetbase.com/doi/abs/10.1201/b18588-4}
\BIBentrySTDinterwordspacing

\bibitem{blumenthal2010launching}
D.~Blumenthal, ``Launching {HIteCH},'' \emph{New England Journal of Medicine},
  vol. 362, no.~5, pp. 382--385, 2010.

\bibitem{charles2014adoption}
D.~Charles, M.~Gabriel, and M.~Furukawa, ``Adoption of electronic health record
  systems among us non-federal acute care hospitals: 2008-2013. 2014,'' 2014.

\bibitem{hsiao2012use}
C.-J. Hsiao, E.~Hing \emph{et~al.}, \emph{Use and Characteristics of Electronic
  Health Record Systems Among Office-Based Physician Practices, United States,
  2001-2013}.\hskip 1em plus 0.5em minus 0.4em\relax US Department of Health
  and Human Services, Centers for Disease Control and Prevention, National
  Center for Health Statistics, 2014.

\bibitem{McCarthy2015}
J.~McCarthy, ``Doctors like {EHRs} even less than they did five years ago,''
  \emph{Healthcare IT News}, 2015.

\bibitem{snell2015}
\BIBentryALTinterwordspacing
E.~Snell, ``Top 10 healthcare data breaches of 2015,'' 2015. [Online].
  Available:
  \url{http://healthitsecurity.com/news/top-10-healthcare-data-breaches-of-2015}
\BIBentrySTDinterwordspacing

\bibitem{clynch2015medical}
N.~Clynch and J.~Kellett, ``Medical documentation: Part of the solution, or
  part of the problem? {A} narrative review of the literature on the time spent
  on and value of medical documentation,'' \emph{International journal of
  medical informatics}, vol.~84, no.~4, pp. 221--228, 2015.

\bibitem{friedberg2013factors}
M.~W. Friedberg, P.~G. Chen, F.~M. Aunon, K.~R. Van~Busum, C.~Pham, J.~P.
  Caloyeras, S.~Mattke, E.~Pitchforth, D.~D. Quigley, R.~H. Brook
  \emph{et~al.}, \emph{Factors affecting physician professional satisfaction
  and their implications for patient care, health systems, and health
  policy}.\hskip 1em plus 0.5em minus 0.4em\relax Rand Corporation, 2013.

\bibitem{mcdonald2014use}
C.~J. McDonald, F.~M. Callaghan, A.~Weissman, R.~M. Goodwin, M.~Mundkur, and
  T.~Kuhn, ``Use of internist's free time by ambulatory care electronic medical
  record systems,'' \emph{JAMA internal medicine}, vol. 174, no.~11, pp.
  1860--1863, 2014.

\bibitem{mcdonald2012invited}
C.~J. McDonald and M.~H. McDonald, ``Invited commentary --- electronic medical
  records and preserving primary care physicians' time,'' \emph{Archives of
  internal medicine}, vol. 172, no.~3, pp. 285--287, 2012.

\bibitem{mihalef2011patient}
V.~Mihalef, R.~I. Ionasec, P.~Sharma, B.~Georgescu, I.~Voigt, M.~Suehling, and
  D.~Comaniciu, ``Patient-specific modelling of whole heart anatomy, dynamics
  and haemodynamics from four-dimensional cardiac {CT} images,''
  \emph{Interface Focus}, vol.~1, no.~3, pp. 286--296, 2011.

\bibitem{DBLP:books/crc/p/PadfieldMG15}
\BIBentryALTinterwordspacing
D.~R. Padfield, P.~R.~S. Mendon{\c{c}}a, and S.~Gupta, ``Biomedical image
  analysis,'' in \emph{Healthcare Data Analytics.}, 2015, pp. 61--89. [Online].
  Available: \url{http://www.crcnetbase.com/doi/abs/10.1201/b18588-5}
\BIBentrySTDinterwordspacing

\bibitem{DBLP:books/crc/p/LiaoYWHZSPZBS15}
\BIBentryALTinterwordspacing
S.~Liao, S.~Yu, M.~Wolf, G.~Hermosillo, Y.~Zhan, Y.~Shinagawa, Z.~Peng, X.~S.
  Zhou, L.~Bogoni, and M.~Salganicoff, ``Computer-assisted medical image
  analysis systems,'' in \emph{Healthcare Data Analytics.}, 2015, pp. 657--683.
  [Online]. Available:
  \url{http://www.crcnetbase.com/doi/abs/10.1201/b18588-24}
\BIBentrySTDinterwordspacing

\bibitem{enlitic2015}
J.~Novet, ``Deep learning startup {Enlitic} raises \$10m from radiology company
  {Capitol Health},'' \emph{venturebeat}, 2015.

\bibitem{tresp2013towards}
V.~Tresp, S.~Zillner, M.~J. Costa, Y.~Huang, A.~Cavallaro, P.~A. Fasching,
  A.~Reis, M.~Sedlmayr, T.~Ganslandt, K.~Budde \emph{et~al.}, ``Towards a new
  science of a clinical data intelligence,'' \emph{arXiv preprint
  arXiv:1311.4180}, 2013.

\bibitem{DBLP:books/crc/p/RajaJ15}
\BIBentryALTinterwordspacing
K.~Raja and S.~Jonnalagadda, ``Natural language processing and data mining for
  clinical text,'' in \emph{Healthcare Data Analytics.}, 2015, pp. 219--249.
  [Online]. Available: \url{http://www.crcnetbase.com/doi/abs/10.1201/b18588-9}
\BIBentrySTDinterwordspacing

\bibitem{mcdonald1998canopy}
C.~J. McDonald, J.~M. Overhage, P.~R. Dexter, L.~Blevins, J.~Meeks-Johnson,
  J.~G. Suico, M.~C. Tucker, and G.~Schadow, ``Canopy computing: using the web
  in clinical practice,'' \emph{Jama}, vol. 280, no.~15, pp. 1325--1329, 1998.

\bibitem{Bresnick2015}
J.~Bresnick, ``How can healthcare big data analytics bust data silos?''
  \emph{Healthcare IT News}, 2015.

\bibitem{murphy2010serving}
S.~N. Murphy, G.~Weber, M.~Mendis, V.~Gainer, H.~C. Chueh, S.~Churchill, and
  I.~Kohane, ``Serving the enterprise and beyond with informatics for
  integrating biology and the bedside (i2b2),'' \emph{Journal of the American
  Medical Informatics Association}, vol.~17, no.~2, pp. 124--130, 2010.

\bibitem{athey2013transmart}
B.~D. Athey, M.~Braxenthaler, M.~Haas, and Y.~Guo, ``{tranSMART}: an open
  source and community-driven informatics and data sharing platform for
  clinical and translational research,'' \emph{AMIA Summits on Translational
  Science Proceedings}, vol. 2013, p.~6, 2013.

\bibitem{DBLP:books/crc/p/ParkG15}
\BIBentryALTinterwordspacing
Y.~Park and J.~Ghosh, ``Privacy-preserving data publishing methods in
  healthcare,'' in \emph{Healthcare Data Analytics.}, 2015, pp. 507--529.
  [Online]. Available:
  \url{http://www.crcnetbase.com/doi/abs/10.1201/b18588-18}
\BIBentrySTDinterwordspacing

\bibitem{danezis2015privacy}
G.~Danezis, J.~Domingo-Ferrer, M.~Hansen, J.-H. Hoepman, D.~L. Metayer,
  R.~Tirtea, and S.~Schiffner, ``Privacy and data protection by design-from
  policy to engineering,'' \emph{arXiv preprint arXiv:1501.03726}, 2015.

\bibitem{raths2015}
D.~Raths, ``{UPMC} funds pittsburgh health data alliance,'' \emph{Health
  Affairs}, 2015.

\bibitem{cmu-spyce}
\BIBentryALTinterwordspacing
B.~Spice, ``The future of health care is in the data,'' 2015. [Online].
  Available: \url{https://www.cs.cmu.edu/news/future-health-care-data}
\BIBentrySTDinterwordspacing

\bibitem{Flanagan2014}
W.~Flanagan, ``{UIUC} and the {Mayo Clinic} get \$9.3 million to try and solve
  the biomed big data puzzle,'' \emph{chicagoinno}, 2014.

\bibitem{marcum-2014}
\BIBentryALTinterwordspacing
C.~Marcum, ``The rise of big data in health care,'' 2014. [Online]. Available:
  \url{http://www.kpihp.org/how-big-data-can-inform-healthcare-decisions/\#sthash.UbfR4h4s.dpbs}
\BIBentrySTDinterwordspacing

\bibitem{Byron2014}
J.~Byron, ``Big data improves care for {Kaiser Permanente’s} smallest
  members,'' \emph{Kaiser Permanente Division of Research}, 2014.

\bibitem{overhage2013desideratum}
J.~M. Overhage, P.~B. Ryan, M.~J. Schuemie, and P.~E. Stang, ``Desideratum for
  evidence based epidemiology,'' \emph{Drug safety}, vol.~36, no.~1, pp. 5--14,
  2013.

\bibitem{hripcsak2015observational}
G.~Hripcsak, J.~Duke, N.~Shah, C.~Reich, V.~Huser, M.~Schuemie, M.~Suchard,
  R.~Park, I.~Wong, P.~Rijnbeek \emph{et~al.}, ``Observational health data
  sciences and informatics ({OHDSI}): opportunities for observational
  researchers,'' \emph{MEDINFO}, vol.~15, 2015.

\bibitem{sonntag2015clinical}
D.~Sonntag, V.~Tresp, S.~Zillner, A.~Cavallaro, M.~Hammon, A.~Reis, P.~A.
  Fasching, M.~Sedlmayr, T.~Ganslandt, H.-U. Prokosch \emph{et~al.}, ``The
  clinical data intelligence project,'' \emph{Informatik-Spektrum}, pp. 1--11,
  2015.

\bibitem{esteban2015}
C.~Esteban, D.~Schmidt, D.~Krompass, and V.~Tresp, ``Predicting sequences of
  clinical events by using a personalized temporal latent embedding model,''
  \emph{IEEE International Conference on Healthcare Informatics (ICHI)}, 2015.

\bibitem{dmh2016}
\BIBentryALTinterwordspacing
J.~Novet, ``Google’s deepmind ai group unveils health care ambitions,'' 2016.
  [Online]. Available:
  \url{http://venturebeat.com/2016/02/24/googles-deepmind-ai-group-unveils-heath-care-ambitions/}
\BIBentrySTDinterwordspacing

\bibitem{esteban2016predicting}
C.~Esteban, O.~Staeck, Y.~Yang, and V.~Tresp, ``Predicting clinical events by
  combining static and dynamic information using recurrent neural networks,''
  \emph{Proceedings of the IEEE International Conference on Healthcare
  Informatics (ICHI)}, 2016.

\bibitem{harvard2014}
\BIBentryALTinterwordspacing
{Harvard Business Review}, ``How big data impacts healthcare,'' 2014. [Online].
  Available:
  \url{https://hbr.org/resources/pdfs/comm/sap/18826-HBR-SAP-Healthcare-Aug-2014.pdf}
\BIBentrySTDinterwordspacing

\bibitem{tomek2012collaborative}
I.~M. Tomek, A.~L. Sabel, M.~I. Froimson, G.~Muschler, D.~S. Jevsevar, K.~M.
  Koenig, D.~G. Lewallen, J.~M. Naessens, L.~A. Savitz, J.~L. Westrich
  \emph{et~al.}, ``A collaborative of leading health systems finds wide
  variations in total knee replacement delivery and takes steps to improve
  value,'' \emph{Health Affairs}, pp. 10--1377, 2012.

\bibitem{Lampe2015}
\BIBentryALTinterwordspacing
D.~Lampe, ``{U-M} launching \$100 million data science initiative,'' 2015.
  [Online]. Available:
  \url{https://record.umich.edu/articles/u-m-launching-100-million-data-science-initiative}
\BIBentrySTDinterwordspacing

\bibitem{Naegele2015}
\BIBentryALTinterwordspacing
D.~Naegele, ``Analytics tool predicts readmission with 82\% accuracy,'' 2015.
  [Online]. Available:
  \url{http://www.infieldhealth.com/blog/analytics-tool-predicts-readmission-with-82-accuracy}
\BIBentrySTDinterwordspacing

\bibitem{Penn2016}
\BIBentryALTinterwordspacing
{Penn Medicine}, ``Penn research team receives \$5 million grant to use big
  data to improve health,'' 2016. [Online]. Available:
  \url{http://www.uphs.upenn.edu/news/News-Releases/2016/02/polsky/}
\BIBentrySTDinterwordspacing

\bibitem{Bresnick2015-3}
J.~Bresnick, ``Precision medicine, big data analytics intersect for better
  care,'' \emph{Healthcare IT Analytics}, 2015.

\bibitem{Forum2015}
{Forum members}, ``Stratified, personalised or {P4} medicine: a new direction
  for placing the patient at the centre of healthcare and health education (may
  2015) summary of a joint forum meeting held on 12 may 2015,'' Supported by
  the Academy of Medical Sciences, the University of Southampton, Science
  Europe and the Medical Research Council, Tech. Rep., 2015.

\bibitem{harris2006use}
A.~D. Harris, J.~C. McGregor, E.~N. Perencevich, J.~P. Furuno, J.~Zhu, D.~E.
  Peterson, and J.~Finkelstein, ``The use and interpretation of
  quasi-experimental studies in medical informatics,'' \emph{Journal of the
  American Medical Informatics Association}, vol.~13, no.~1, pp. 16--23, 2006.

\bibitem{Rubenfirel2013}
A.~Rubenfire, ``Hospitals use big-data platform to improve care,'' \emph{Modern
  Healthcare}, 2015.

\bibitem{huang2016path}
B.~E. Huang, W.~Mulyasasmita, and G.~Rajagopal, ``The path from big data to
  precision medicine,'' \emph{Expert Review of Precision Medicine and Drug
  Development}, no. just-accepted, 2016.

\bibitem{fan2014challenges}
J.~Fan, F.~Han, and H.~Liu, ``Challenges of big data analysis,'' \emph{National
  science review}, vol.~1, no.~2, pp. 293--314, 2014.

\bibitem{dahl2008data}
F.~A. Dahl, M.~Grotle, J.~{\v{S}}. Benth, and B.~Natvig, ``Data splitting as a
  countermeasure against hypothesis fishing: with a case study of predictors
  for low back pain,'' \emph{European journal of epidemiology}, vol.~23, no.~4,
  pp. 237--242, 2008.

\bibitem{tang2007comparison}
P.~C. Tang, M.~Ralston, M.~F. Arrigotti, L.~Qureshi, and J.~Graham,
  ``Comparison of methodologies for calculating quality measures based on
  administrative data versus clinical data from an electronic health record
  system: implications for performance measures,'' \emph{Journal of the
  American Medical Informatics Association}, vol.~14, no.~1, pp. 10--15, 2007.

\bibitem{rosenman2013agreement}
M.~Rosenman, J.~He, J.~Martin, K.~Nutakki, I.~Gradus-Pizlo, and S.~L. Hui,
  ``Agreement between claims and electronic medical records data for {CHF} in
  inpatients,'' in \emph{Pharmacoepidemiology And Drug Safety}, vol.~22, 2013,
  pp. 269--269.

\bibitem{chandola2015fraud}
\BIBentryALTinterwordspacing
V.~Chandola, J.~C. Schryver, and S.~R. Sukumar, ``Fraud detection in
  healthcare,'' in \emph{Healthcare Data Analytics}, 2015, pp. 577--598.
  [Online]. Available:
  \url{http://www.crcnetbase.com/doi/abs/10.1201/b18588-21}
\BIBentrySTDinterwordspacing

\bibitem{Sullivan2009}
K.~M. Sullivan, ``But doctor, {I} still have both feet! {Remedial} problems
  faced by victims of medical identity theft,'' \emph{Am J Law Med.}, vol.~35,
  no.~4, pp. 647--81, 2009.

\bibitem{betz2012experiences}
A.~Betz, ``The experiences of adult/child identity theft victims,''
  \emph{Digital Repository \@ Iowa State University}, 2012.

\bibitem{chapman2013}
S.~Chapman, ``Capturing cancer data in real time,'' \emph{For The Record},
  2013.

\bibitem{dixon2010framework}
B.~E. Dixon, A.~Zafar, and J.~M. Overhage, ``A framework for evaluating the
  costs, effort, and value of nationwide health information exchange,''
  \emph{Journal of the American Medical Informatics Association}, vol.~17,
  no.~3, pp. 295--301, 2010.

\bibitem{ONC2015}
{https://www.healthit.gov/sites/default/files/nationwide-interoperability-roadmap-draft-version-1.0.pdf},
  ``Connecting health and care for the nation: A shared nationwide
  interoperability roadmap,'' \emph{Office of the National Coordinator for
  Health Information Technology (ONC)}, 2015.

\bibitem{wikiHSC}
\BIBentryALTinterwordspacing
Wikipedia, ``Health and social care information centre --- {W}ikipedia{,} the
  free encyclopedia,'' 2016. [Online]. Available:
  \url{https://en.wikipedia.org/wiki/Health-and-Social-Care-Information-Centre}
\BIBentrySTDinterwordspacing

\bibitem{oecd2015}
\BIBentryALTinterwordspacing
{OECD}, ``Health at a glance 2015,'' \emph{OECD Publishing}, 2015. [Online].
  Available: \url{/content/book/health\_glance-2015-en}
\BIBentrySTDinterwordspacing

\bibitem{mckethan2009improving}
A.~McKethan and B.~P. Center, \emph{Improving Quality and Value in the {US}:
  Health Care System}.\hskip 1em plus 0.5em minus 0.4em\relax Bipartisan Policy
  Center, 2009.

\bibitem{bayer2015public}
R.~Bayer and S.~Galea, ``Public health in the precision-medicine era,''
  \emph{New England Journal of Medicine}, vol. 373, no.~6, pp. 499--501, 2015.

\bibitem{Bresnick2015-4}
J.~Bresnick, ``Is there conflict between precision medicine, population
  health?'' \emph{Healthcare IT Analytics}, 2015.

\bibitem{onlinetech2015}
{Multiple authors}, ``{HIPAA} compliant hosting,'' \emph{OnLINE TECH}, 2015.

\bibitem{tucker2014}
M.~E. Tucker, ``Doctors, not just patients, use {Wikipedia}, too: {IMS}
  report,'' \emph{Medscape Medical News}, 2014.

\bibitem{aoh}
\BIBentryALTinterwordspacing
K.~Christensen, ``The quest for the {Amazon} of healthcare,'' Forbes India,
  Tech. Rep., 2016. [Online]. Available:
  \url{http://forbesindia.com/article/rotman/the-quest-for-the-amazon-of-healthcare}
\BIBentrySTDinterwordspacing

\bibitem{DBLP:books/crc/p/Kotov15}
\BIBentryALTinterwordspacing
A.~Kotov, ``Social media analytics for healthcare,'' in \emph{Healthcare Data
  Analytics.}, 2015, pp. 309--340. [Online]. Available:
  \url{http://www.crcnetbase.com/doi/abs/10.1201/b18588-11}
\BIBentrySTDinterwordspacing

\bibitem{horvitz2015data}
E.~Horvitz and D.~Mulligan, ``Data, privacy, and the greater good,''
  \emph{Science}, vol. 349, no. 6245, pp. 253--255, 2015.

\bibitem{lazer2014parable}
D.~Lazer, R.~Kennedy, G.~King, and A.~Vespignani, ``The parable of {Google}
  flu: traps in big data analysis,'' \emph{Science}, vol. 343, no. 14 March,
  2014.

\bibitem{olson2013reassessing}
D.~R. Olson, K.~J. Konty, M.~Paladini, C.~Viboud, and L.~Simonsen,
  ``Reassessing {Google} flu trends data for detection of seasonal and pandemic
  influenza: a comparative epidemiological study at three geographic scales,''
  \emph{PLoS Comput Biol}, vol.~9, no.~10, p. e1003256, 2013.

\bibitem{janssens2012research}
A.~C.~J. Janssens and P.~Kraft, ``Research conducted using data obtained
  through online communities: ethical implications of methodological
  limitations,'' \emph{PLOS}, 2012.

\bibitem{kalf2013predictive}
R.~Kalf, R.~Bakker, and C.~Janssens, ``Predictive ability of direct-to-consumer
  pharmacogenetic testing: when is lack of evidence really lack of evidence?''
  \emph{Pharmacogenomics}, vol.~14, no.~4, pp. 341--344, 2013.

\bibitem{Wilkinson2016}
J.~Wilkinson, ``How companies are secretly tracking employees' health and
  private lives with 'big data' to save money,'' \emph{DailyMail}, 2016.

\bibitem{gupta2011patientslikeme}
S.~Gupta and J.~Riis, ``Patientslikeme: An online community of patients,''
  \emph{Harvard Business School Marketing Unit Case}, no. 511-093, 2011.

\bibitem{brownstein2009power}
C.~A. Brownstein, J.~S. Brownstein, D.~S. Williams, P.~Wicks, and J.~A.
  Heywood, ``The power of social networking in medicine,'' \emph{Nature
  biotechnology}, vol.~27, no.~10, pp. 888--890, 2009.

\bibitem{frost2008social}
J.~H. Frost and M.~P. Massagli, ``Social uses of personal health information
  within patientslikeme, an online patient community: what can happen when
  patients have access to one another’s data,'' \emph{Journal of Medical
  Internet Research}, vol.~10, no.~3, 2008.

\bibitem{wicks2010sharing}
P.~Wicks, M.~Massagli, J.~Frost, C.~Brownstein, S.~Okun, T.~Vaughan,
  R.~Bradley, and J.~Heywood, ``Sharing health data for better outcomes on
  patientslikeme,'' \emph{Journal of medical Internet research}, vol.~12,
  no.~2, 2010.

\bibitem{sunyaev2010evaluation}
A.~Sunyaev, D.~Chornyi, C.~Mauro, and H.~Krcmar, ``Evaluation framework for
  personal health records: {Microsoft HealthVault} vs. {Google} health,'' in
  \emph{System Sciences (HICSS), 2010 43rd Hawaii International Conference
  on}.\hskip 1em plus 0.5em minus 0.4em\relax IEEE, 2010, pp. 1--10.

\bibitem{collins2012fulfill}
F.~Collins, ``How to fulfill the true promise of “mhealth”,''
  \emph{Scientific American}, vol. 307, no.~1, pp. 16--16, 2012.

\bibitem{kay2011mhealth}
M.~Kay, J.~Santos, and M.~Takane, ``mhealth: New horizons for health through
  mobile technologies,'' \emph{World Health Organization}, pp. 66--71, 2011.

\bibitem{Fox2016}
\BIBentryALTinterwordspacing
S.~Fox and M.~Duggan, ``Main findings: Mobile health,'' 2016. [Online].
  Available: \url{http://www.pewinternet.org/2012/11/08/main-findings-6/}
\BIBentrySTDinterwordspacing

\bibitem{gagnon2016m}
M.-P. Gagnon, P.~Ngangue, J.~Payne-Gagnon, and M.~Desmartis, ``m-health
  adoption by healthcare professionals: a systematic review,'' \emph{Journal of
  the American Medical Informatics Association}, vol.~23, no.~1, pp. 212--220,
  2016.

\bibitem{free2013effectiveness}
C.~Free, G.~Phillips, L.~Watson, L.~Galli, L.~Felix, P.~Edwards, V.~Patel, and
  A.~Haines, ``The effectiveness of mobile-health technologies to improve
  health care service delivery processes: a systematic review and
  meta-analysis,'' \emph{PLoS Med}, vol.~10, no.~1, p. e1001363, 2013.

\bibitem{aranda2014systematic}
C.~B. Aranda-Jan, N.~Mohutsiwa-Dibe, and S.~Loukanova, ``Systematic review on
  what works, what does not work and why of implementation of mobile health
  (mhealth) projects in {Africa},'' \emph{BMC public health}, vol.~14, no.~1,
  p. 188, 2014.

\bibitem{miyamoto2016tracking}
S.~W. Miyamoto, S.~Henderson, H.~M. Young, A.~Pande, and J.~J. Han, ``Tracking
  health data is not enough: A qualitative exploration of the role of
  healthcare partnerships and mhealth technology to promote physical activity
  and to sustain behavior change,'' \emph{JMIR mHealth and uHealth}, vol.~4,
  no.~1, p.~e5, 2016.

\bibitem{jovanov2011body}
E.~Jovanov and A.~Milenkovic, ``Body area networks for ubiquitous healthcare
  applications: opportunities and challenges,'' \emph{Journal of medical
  systems}, vol.~35, no.~5, pp. 1245--1254, 2011.

\bibitem{laksanasopin2015smartphone}
T.~Laksanasopin, T.~W. Guo, S.~Nayak, A.~A. Sridhara, S.~Xie, O.~O. Olowookere,
  P.~Cadinu, F.~Meng, N.~H. Chee, J.~Kim \emph{et~al.}, ``A smartphone dongle
  for diagnosis of infectious diseases at the point of care,'' \emph{Science
  translational medicine}, vol.~7, no. 273, pp. 273re1--273re1, 2015.

\bibitem{knowlton2015sickle}
S.~Knowlton, I.~Sencan, Y.~Aytar, J.~Khoory, M.~Heeney, I.~Ghiran, and
  S.~Tasoglu, ``Sickle cell detection using a smartphone,'' \emph{Scientific
  reports}, vol.~5, 2015.

\bibitem{burke2015current}
L.~E. Burke, J.~Ma, K.~M. Azar, G.~G. Bennett, E.~D. Peterson, Y.~Zheng,
  W.~Riley, J.~Stephens, S.~H. Shah, B.~Suffoletto \emph{et~al.}, ``Current
  science on consumer use of mobile health for cardiovascular disease
  prevention a scientific statement from the american heart association,''
  \emph{Circulation}, vol. 132, no.~12, pp. 1157--1213, 2015.

\bibitem{hamine2015impact}
S.~Hamine, E.~Gerth-Guyette, D.~Faulx, B.~B. Green, and A.~S. Ginsburg,
  ``Impact of mhealth chronic disease management on treatment adherence and
  patient outcomes: a systematic review,'' \emph{Journal of medical Internet
  research}, vol.~17, no.~2, 2015.

\bibitem{Crotti2015}
N.~Crotti, ``How the {Apple} watch can collect patient data,'' \emph{EE Times},
  2015.

\bibitem{Herper2015}
M.~Herper, ``Can {Apple} and {IBM} change health care? {Five} big questions,''
  \emph{Forbes}, 2015.

\bibitem{swan2013quantified}
M.~Swan, ``The quantified self: Fundamental disruption in big data science and
  biological discovery,'' \emph{Big Data}, vol.~1, no.~2, pp. 85--99, 2013.

\bibitem{Heather2015}
\BIBentryALTinterwordspacing
H.~Caouette, ``Harris poll survey finds patients want a deeper digital
  connection with their doctors,'' 2015. [Online]. Available:
  \url{https://www.eclinicalworks.com/pr-harris-poll-patient-engagement-survey/}
\BIBentrySTDinterwordspacing

\bibitem{hamel2014fda}
M.~B. Hamel, N.~G. Cortez, I.~G. Cohen, and A.~S. Kesselheim, ``{FDA}
  regulation of mobile health technologies,'' \emph{New England Journal of
  Medicine}, vol. 371, no.~4, pp. 372--379, 2014.

\bibitem{ogino2013molecular}
S.~Ogino, P.~Lochhead, A.~T. Chan, R.~Nishihara, E.~Cho, B.~M. Wolpin, J.~A.
  Meyerhardt, A.~Meissner, E.~S. Schernhammer, C.~S. Fuchs \emph{et~al.},
  ``Molecular pathological epidemiology of epigenetics: emerging integrative
  science to analyze environment, host, and disease,'' \emph{Modern Pathology},
  vol.~26, no.~4, pp. 465--484, 2013.

\bibitem{lander2001initial}
E.~S. Lander, L.~M. Linton, B.~Birren, C.~Nusbaum, M.~C. Zody, J.~Baldwin,
  K.~Devon, K.~Dewar, M.~Doyle, W.~FitzHugh \emph{et~al.}, ``Initial sequencing
  and analysis of the human genome,'' \emph{Nature}, vol. 409, no. 6822, pp.
  860--921, 2001.

\bibitem{venter2001sequence}
J.~C. Venter, M.~D. Adams, E.~W. Myers, P.~W. Li, R.~J. Mural, G.~G. Sutton,
  H.~O. Smith, M.~Yandell, C.~A. Evans, R.~A. Holt \emph{et~al.}, ``The
  sequence of the human genome,'' \emph{Science}, vol. 291, no. 5507, pp.
  1304--1351, 2001.

\bibitem{kulkarni2013personalized}
S.~Kulkarni and P.~Ma, ``Personalized medicine: The path forward,''
  \emph{McKinsey \& Company}, vol.~3, pp. 1--48, 2013.

\bibitem{duncan2014revolution}
E.~Duncan, M.~Brown, and E.~M. Shore, ``The revolution in human monogenic
  disease mapping,'' \emph{Genes}, vol.~5, no.~3, pp. 792--803, 2014.

\bibitem{anand2008cancer}
P.~Anand, A.~B. Kunnumakara, C.~Sundaram, K.~B. Harikumar, S.~T. Tharakan,
  O.~S. Lai, B.~Sung, and B.~B. Aggarwal, ``Cancer is a preventable disease
  that requires major lifestyle changes,'' \emph{Pharmaceutical research},
  vol.~25, no.~9, pp. 2097--2116, 2008.

\bibitem{visscher2012five}
P.~M. Visscher, M.~A. Brown, M.~I. McCarthy, and J.~Yang, ``Five years of
  {GWAS} discovery,'' \emph{The American Journal of Human Genetics}, vol.~90,
  no.~1, pp. 7--24, 2012.

\bibitem{cancer2012comprehensive}
{Cancer Genome Atlas Network and others}, ``Comprehensive molecular portraits
  of human breast tumours,'' \emph{Nature}, vol. 490, no. 7418, pp. 61--70,
  2012.

\bibitem{bettegowda2014detection}
C.~Bettegowda, M.~Sausen, R.~J. Leary, I.~Kinde, Y.~Wang, N.~Agrawal, B.~R.
  Bartlett, H.~Wang, B.~Luber, R.~M. Alani \emph{et~al.}, ``Detection of
  circulating tumor {DNA} in early-and late-stage human malignancies,''
  \emph{Science translational medicine}, vol.~6, no. 224, pp. 224ra24--224ra24,
  2014.

\bibitem{nevins2007mining}
J.~R. Nevins and A.~Potti, ``Mining gene expression profiles: expression
  signatures as cancer phenotypes,'' \emph{Nature Reviews Genetics}, vol.~8,
  no.~8, pp. 601--609, 2007.

\bibitem{catenacci2015mass}
D.~V. Catenacci, W.-L. Liao, L.~Zhao, E.~Whitcomb, L.~Henderson, E.~O’Day,
  P.~Xu, S.~Thyparambil, D.~Krizman, K.~Bengali \emph{et~al.},
  ``Mass-spectrometry-based quantitation of {Her2} in gastroesophageal tumor
  tissue: comparison to {IHC} and {FISH},'' \emph{Gastric Cancer}, pp. 1--14,
  2015.

\bibitem{Tannock2016}
I.~F. Tannock and J.~A. Hickman, ``Limits to personalized cancer medicine,''
  \emph{New England Journal of Medicine}, vol. 375, 2016.

\bibitem{miki1994strong}
Y.~Miki, J.~Swensen, D.~Shattuck-Eidens, P.~A. Futreal, K.~Harshman,
  S.~Tavtigian, Q.~Liu, C.~Cochran, L.~M. Bennett, W.~Ding \emph{et~al.}, ``A
  strong candidate for the breast and ovarian cancer susceptibility gene
  {BRCA1},'' \emph{Science}, vol. 266, no. 5182, pp. 66--71, 1994.

\bibitem{wooster1995identification}
R.~Wooster, G.~Bignell, J.~Lancaster, S.~Swift, S.~Seal, J.~Mangion,
  N.~Collins, S.~Gregory, C.~Gumbs, G.~Micklem \emph{et~al.}, ``Identification
  of the breast cancer susceptibility gene {BRCA2},'' \emph{Nature}, vol. 378,
  no. 6559, pp. 789--792, 1995.

\bibitem{slamon1987human}
D.~J. Slamon, G.~M. Clark, S.~G. Wong, W.~J. Levin, A.~Ullrich, and W.~L.
  McGuire, ``Human breast cancer: correlation of relapse and survival with
  amplification of the {HER-2/neu} oncogene,'' \emph{Science}, vol. 235, no.
  4785, pp. 177--182, 1987.

\bibitem{spear2001clinical}
B.~B. Spear, M.~Heath-Chiozzi, and J.~Huff, ``Clinical application of
  pharmacogenetics,'' \emph{Trends in molecular medicine}, vol.~7, no.~5, pp.
  201--204, 2001.

\bibitem{rastegar2015toward}
M.~Rastegar-Mojarad and R.~Prasad, ``Toward a complete database of drug
  repurposing candidates extracted from social media, biomedical literature,
  and genetic data,'' in \emph{Healthcare Informatics (ICHI), 2015
  International Conference on}.\hskip 1em plus 0.5em minus 0.4em\relax IEEE,
  2015, pp. 494--494.

\bibitem{aliper2016deep}
A.~Aliper, S.~Plis, A.~Artemov, A.~Ulloa, P.~Mamoshina, and A.~Zhavoronkov,
  ``Deep learning applications for predicting pharmacological properties of
  drugs and drug repurposing using transcriptomic data,'' \emph{Molecular
  pharmaceutics}, 2016.

\bibitem{hyman2015vemurafenib}
D.~M. Hyman, I.~Puzanov, V.~Subbiah, J.~E. Faris, I.~Chau, J.-Y. Blay, J.~Wolf,
  N.~S. Raje, E.~L. Diamond, A.~Hollebecque \emph{et~al.}, ``Vemurafenib in
  multiple nonmelanoma cancers with braf v600 mutations,'' \emph{New England
  Journal of Medicine}, vol. 373, no.~8, pp. 726--736, 2015.

\bibitem{green2011charting}
E.~D. Green, M.~S. Guyer, N.~H. G.~R. Institute \emph{et~al.}, ``Charting a
  course for genomic medicine from base pairs to bedside,'' \emph{Nature}, vol.
  470, no. 7333, pp. 204--213, 2011.

\bibitem{fackler2009paving}
J.~L. Fackler and A.~L. McGuire, ``Paving the way to personalized genomic
  medicine: steps to successful implementation,'' \emph{Current
  pharmacogenomics and personalized medicine}, vol.~7, no.~2, p. 125, 2009.

\bibitem{michailidou2013large}
K.~Michailidou, P.~Hall, A.~Gonzalez-Neira, M.~Ghoussaini, J.~Dennis, R.~L.
  Milne, M.~K. Schmidt, J.~Chang-Claude, S.~E. Bojesen, M.~K. Bolla
  \emph{et~al.}, ``Large-scale genotyping identifies 41 new loci associated
  with breast cancer risk,'' \emph{Nature genetics}, vol.~45, no.~4, pp.
  353--361, 2013.

\bibitem{10002010map}
{1000 Genomes Project Consortium and others}, ``A map of human genome variation
  from population-scale sequencing,'' \emph{Nature}, vol. 467, no. 7319, pp.
  1061--1073, 2010.

\bibitem{hirtzlin2003empirical}
I.~Hirtzlin, C.~Dubreuil, N.~Pr{\'e}aubert, J.~Duchier, B.~Jansen, J.~Simon,
  P.~L. de~Faria, A.~Perez-Lezaun, B.~Visser, G.~D. Williams \emph{et~al.},
  ``An empirical survey on biobanking of human genetic material and data in six
  {EU} countries,'' \emph{European Journal of Human Genetics}, vol.~11, no.~6,
  pp. 475--488, 2003.

\bibitem{soon2016abstract}
P.~Soon-Shiong, S.~Rabizadeh, S.~Benz, F.~Cecchi, T.~Hembrough, E.~Mahen,
  K.~Burton, C.~Song, F.~Senecal, S.~Schmechel \emph{et~al.}, ``Abstract
  p6-05-08: Integrating whole exome sequencing data with {RNAseq} and
  quantitative proteomics to better inform clinical treatment decisions in
  patients with metastatic triple negative breast cancer,'' \emph{Cancer
  Research}, vol.~76, no. 4 Supplement, pp. P6--05, 2016.

\bibitem{Hayden2015}
C.~Hayden, ``Genome researchers raise alarm over big data,'' \emph{Nature
  News}, 2015.

\bibitem{stephens2015big}
Z.~D. Stephens, S.~Y. Lee, F.~Faghri, R.~H. Campbell, C.~Zhai, M.~J. Efron,
  R.~Iyer, M.~C. Schatz, S.~Sinha, and G.~E. Robinson, ``Big data: Astronomical
  or genomical?'' \emph{PLoS Biol}, vol.~13, no.~7, p. e1002195, 2015.

\bibitem{shirts2015cser}
B.~H. Shirts, J.~S. Salama, S.~J. Aronson, W.~K. Chung, S.~W. Gray, L.~A.
  Hindorff, G.~P. Jarvik, S.~E. Plon, E.~M. Stoffel, P.~Z. Tarczy-Hornoch
  \emph{et~al.}, ``{CSER} and {eMERGE}: current and potential state of the
  display of genetic information in the electronic health record,''
  \emph{Journal of the American Medical Informatics Association}, p. ocv065,
  2015.

\bibitem{shah2016building}
A.~Shah, A.~K. Stewart, A.~Kolacevski, D.~Michels, and R.~Miller, ``Building a
  rapid learning health care system for oncology: Why {CancerLinQ} collects
  identifiable health information to achieve its vision,'' \emph{Journal of
  Clinical Oncology}, p. JCO650598, 2016.

\bibitem{abernethy2013asco}
A.~Abernethy, ``{ASCO's CancerLinQ} and breast cancer outcomes,'' in
  \emph{European Journal Of Cancer}, vol.~49.\hskip 1em plus 0.5em minus
  0.4em\relax Elsevier Sci Ltd The Boulevard, Langford Lane, Kidlington, Oxford
  Ox5 1Gb, Oxon, England, 2013, pp. S37--S37.

\bibitem{ratner2015ibm}
M.~Ratner, ``{IBM}'s {Watson} group signs up genomics partners,'' \emph{Nature
  biotechnology}, vol.~33, no.~1, pp. 10--11, 2015.

\bibitem{savvy2015watson}
{Tech Savvy}, ``Watson will see you now: a supercomputer to help clinicians
  make informed treatment decisions,'' 2015.

\bibitem{tureci2016targeting}
{\"O}.~T{\"u}reci, M.~Vormehr, M.~Diken, S.~Kreiter, C.~Huber, and U.~Sahin,
  ``Targeting the heterogeneity of cancer with individualized neoepitope
  vaccines,'' \emph{Clinical Cancer Research}, vol.~22, no.~8, pp. 1885--1896,
  2016.

\bibitem{srivastava2015neoepitopes}
P.~K. Srivastava, ``Neoepitopes of cancers: Looking back, looking ahead,''
  \emph{Cancer immunology research}, vol.~3, no.~9, pp. 969--977, 2015.

\bibitem{jones2015personalized}
S.~Jones, V.~Anagnostou, K.~Lytle, S.~Parpart-Li, M.~Nesselbush, D.~R. Riley,
  M.~Shukla, B.~Chesnick, M.~Kadan, E.~Papp \emph{et~al.}, ``Personalized
  genomic analyses for cancer mutation discovery and interpretation,''
  \emph{Science translational medicine}, vol.~7, no. 283, pp. 283ra53--283ra53,
  2015.

\bibitem{morse2013novel}
M.~A. Morse, A.~Chaudhry, E.~S. Gabitzsch, A.~C. Hobeika, T.~Osada, T.~M. Clay,
  A.~Amalfitano, B.~K. Burnett, G.~R. Devi, D.~S. Hsu \emph{et~al.}, ``Novel
  adenoviral vector induces {T-cell} responses despite anti-adenoviral
  neutralizing antibodies in colorectal cancer patients,'' \emph{Cancer
  Immunology, Immunotherapy}, vol.~62, no.~8, pp. 1293--1301, 2013.

\bibitem{balint2015extended}
J.~P. Balint, E.~S. Gabitzsch, A.~Rice, Y.~Latchman, Y.~Xu, G.~L.
  Messerschmidt, A.~Chaudhry, M.~A. Morse, and F.~R. Jones, ``Extended
  evaluation of a phase 1/2 trial on dosing, safety, immunogenicity, and
  overall survival after immunizations with an advanced-generation {Ad5 [E1-,
  E2b-]-CEA (6D)} vaccine in late-stage colorectal cancer,'' \emph{Cancer
  Immunology, Immunotherapy}, vol.~64, no.~8, pp. 977--987, 2015.

\bibitem{collins2015new}
F.~S. Collins and H.~Varmus, ``A new initiative on precision medicine,''
  \emph{New England Journal of Medicine}, vol. 372, no.~9, pp. 793--795, 2015.

\end{thebibliography}

\end{document}